\documentclass{article}

\usepackage{fullpage}
\usepackage{graphicx}
\graphicspath{{.}}

\usepackage{hyperref}
\usepackage{amssymb}

\usepackage{tikz} 
\usetikzlibrary{backgrounds}
\usetikzlibrary{calc}

\usepackage{natbib}
\bibpunct{(}{)}{;}{a}{}{,}

\definecolor{grey}{rgb}{0.5,0.5,0.5}
\definecolor{lightgrey}{rgb}{0.7,0.7,0.7}
\definecolor{verylightgrey}{rgb}{0.9,0.9,0.9}
\definecolor{darkgrey}{rgb}{0.2,0.2,0.2}

\definecolor{grey5}{rgb}{0.9,0.9,0.9}
\definecolor{grey4}{rgb}{0.833,0.833,0.833}
\definecolor{grey3}{rgb}{0.766,0.766,0.766}
\definecolor{grey2}{rgb}{0.7,0.7,0.7}
\definecolor{grey1}{rgb}{0.533,0.533,0.533}

\title{Micro-level network dynamics of scientific collaboration and impact: relational hyperevent models for the analysis of coauthor networks}

 \author{J\"urgen Lerner\\
   University of Konstanz, Germany\\
   \texttt{juergen.lerner@uni-konstanz.de}
         \and
         Marian-Gabriel H\^{a}ncean\\
         University of Bucharest, Romania\\
         \texttt{gabriel.hancean@sas.unibuc.ro}
 }

 \date{}

\begin{document}

\maketitle

\begin{abstract}
  We discuss a recently proposed family of statistical network models -- relational hyperevent models (RHEM) -- for analyzing team selection and team performance in scientific coauthor networks. The underlying rationale for using RHEM in studies of coauthor networks is that scientific collaboration is intrinsically polyadic, that is, it typically involves teams of any size. Consequently, RHEM specify publication rates associated with hyperedges representing groups of scientists of any size. Going beyond previous work on RHEM for meeting data, we adapt this model family to settings in which relational hyperevents have a dedicated outcome, such as a scientific paper with a measurable impact (e.\,g., the received number of citations). Relational outcome can on the one hand be used to specify additional explanatory variables in RHEM since the probability of coauthoring may be influenced, for instance, by prior (shared) success of scientists. On the other hand relational outcome can also serve as a response variable in models seeking to explain the performance of scientific teams. To tackle the latter we propose relational hyperevent outcome models (RHOM) that are closely related with RHEM to the point that both model families can specify the likelihood of scientific collaboration -- and the expected performance, respectively -- with the same set of explanatory variables allowing to assess, for instance, whether variables leading to increased collaboration also tend to increase scientific impact. For illustration, we apply RHEM to empirical coauthor networks comprising more than 350,000 published papers by scientists working in three scientific disciplines. Our models explain scientific collaboration and impact by, among others, individual activity (preferential attachment), shared activity (familiarity), triadic closure, prior individual and shared success, and prior success disparity among the members of hyperedges. 
\end{abstract}

\section{Introduction}
\label{sec:intro}

Scientific research is increasingly done by teams and papers coauthored by teams are found to have higher impact \citep{wuchty2007increasing}. Such empirical results have stipulated a \emph{science of team science} \citep{borner2010multi} studying the dynamics of scientific collaboration and the output (e.\,g., number of published papers) and impact (e.\,g., number of citations gathered by published papers) of scientific teams \citep{guimera2005team,ahmadpoor2019decoding,hayat2020differential}. Coauthor networks provide observable data on large-scale scientific collaboration and are frequently analyzed to uncover collaboration structure and scientific output and impact \citep{kronegger2012collaboration,ferligoj2015scientific,kumar2015co,hancean2016homophily,cugmas2019scientific}. 

From a modeling perspective, empirical data on scientific collaboration has three distinctive characteristics that we seek to tackle simultaneously with the models proposed in this paper. First, social relations in coauthor networks are given by relational events \citep{b-refsa-08}, rather than by relational states that have inertia. The observations in coauthor network data -- that is, published scientific papers -- are given by interaction events that have associated time stamps. The time ordering is meaningful when analyzing coauthor networks since collaboration probabilities, or the expected impact of collaboration, might depend, among others, on previous publication activity, shared activity, or prior success. A second characteristic is that interaction events in coauthor networks are intrinsically \emph{polyadic} \citep{chodrow2019configuration,chodrow2020annotated}, that is, they involve sets of actors of any size, rather than relating exactly two actors at a time. Ignoring the multi-actor aspect of coauthoring assumes potentially invalid independence of dyads and can lead to information loss. For instance, a triad of three pairwise connected actors might result from a joint three-author publication, or it might result from three different papers, each coauthored by a different dyad but none of them coauthored by all three actors together. Focusing exclusively on dyadic interaction would miss this difference. The combination of time ordering with the multi-actor aspect of publication events makes it challenging to analyze coauthor networks with more common statistical network models such as exponential random graph models (ERGM) \citep{lkr-ergmsn-13}, stochastic actor-oriented models (SAOM) \citep{s-mlnd-05}, or relational event models (REM) \citep{b-refsa-08} but calls for an analysis via RHEM \citep{lerner2021dynamic} which are models for time-stamped multi-actor events. RHEM allow to test, or to control for, higher-order dependencies in coauthor networks which cannot be taken into account in dyadic REM. Third, publication events have a dedicated \emph{outcome} -- the published paper -- which has an observable performance measure, for instance, scientific impact measured via the number of citations received by a paper. This allows not only to model the likelihood of scientific collaboration but also the performance, or expected impact of the work jointly produced by a team of scientists.

The relational outcome associated with hyperevents in coauthor networks calls for extending RHEM that have been defined for meeting events in \citet{lerner2021dynamic} in two directions. First, the impact of past publications can have an influence on the likelihood of future collaborations as scientists might have a preference to collaborate with successful others, or with others with whom they have a history of prior \emph{shared} success. In this direction, we stay within the RHEM framework proposed by \citet{lerner2021dynamic} -- modeling event intensities associated with groups of actors of any size -- but extend the set of explanatory variables to also capture the outcome of past events. Secondly, the relational outcome can be modeled as a response variable in the newly defined RHOM which allow to assess which characteristics of a group of scientists -- such as past activity, familiarity, diversity, or prior success -- tend to increase or decrease future impact of that team.

In this paper we elaborate and apply RHEM \citep{lerner2019rem,lerner2021dynamic} as a general statistical model for scientific collaboration in coauthor networks. RHEM can explain publication rates associated with hyperedges, that is, with subsets of any size from a given population of scientists. Moreover, we propose RHOM \citep{lerner2019rem} as a general model for assessing the determinants of scientific impact of published papers. For illustration we apply RHEM and RHOM to empirical coauthor networks comprising more than 350,000 papers published by EU-based researchers working in three scientific disciplines. We test hypothetical network effects explaining publication rates associated with hyperedges -- as well as the impact of published papers -- by previous publication events in the network. Examples of such network effects include preferential attachment predicting that scientists are inclined to coauthor with others who have already published many papers \citep{kronegger2012collaboration}, triadic closure predicting that scientists are more likely to coauthor if they have previously coauthored with common third actors \citep{ferligoj2015scientific}, or effects explaining co-publication by prior individual success, prior shared success, or prior success disparity \citep{ahmadpoor2019decoding,mukherjee2019prior}.

RHEM and RHOM have some characteristics in common -- but are also distinct in other aspects. Both model frameworks explain response variables associated with hyperedges and thus can be specified with the same set of explanatory variables (\emph{hyperedge statistics}, formally defined in Sect.~\ref{sec:statistics}). From a high-level view, the most crucial differences between RHEM and RHOM are in the set of instances on which they make predictions and in the nature of the response variable. RHEM make predictions for all hyperedges from a given \emph{risk set} -- which can for instance comprise all subsets of a given set of scientists -- and for each of these hyperedges RHEM can specify a separate event rate, or publication intensity. Note that hyperedges in the risk set are not only those that do eventually experience a common event but also those that could potentially have experienced an event, but did not. RHOM on the other hand make predictions for those hyperedges that do experience an event (i.\,e., those sets of scientists that co-publish a paper) and specify the expected outcome of this publication event -- such as the number of citations gathered by a paper -- which can be quantified in a binary (success vs.\ failure), ordinal, or numeric response variable. The fact that RHEM and RHOM can specify collaboration probabilities, or scientific impact respectively, via the same set of explanatory variables allows to assess consistency in team formation and team performance by analyzing whether those predictors that increase co-publication rates (i.\,e., explain team formation) do typically also increase team performance, and \emph{vice versa}.

For clarity of exposition we focus the description in this paper on coauthor networks. However, we note that RHEM and RHOM are not restricted to this application area but can be applied to other situations in which relational hyperevents represent a team that tackles a given task, provides a service, or produces a product and where these events are associated with a measurable outcome. Besides scientific collaboration, other potential application areas include project teams in companies developing a new product or registering a patent \citep{trajtenberg1990penny}, teams of artists and other staff producing a movie \citep{ravasz2003hierarchical}, sports teams \citep{mukherjee2019prior}, or medical teams performing a given surgery \citep{pallotti2020lost}.

In the next section we recall hypergraphs and relational hyperevents and discuss the insufficiency of common statistical network models for hyperevents in coauthor networks. Section~\ref{sec:model} adapts RHEM, proposed in \citep{lerner2021dynamic} for meeting events, to coauthor networks in which hyperevents have an associated relational outcome and proposes RHOM as a model framework seeking to explain outcome. Section~\ref{sec:empirical} illustrates the empirical value of RHEM/RHOM by analyzing an empirical dataset comprising more than 350,000 publication events in three disciplines. In Sect.~\ref{sec:conclusion} we provide concluding remarks and discuss limitations and future work. The appendix presents model variants in which we include additional explanatory variables, analyze interaction effects, and test variations of hyperedge statistics.

\section{Background}
\label{sec:background}

\subsection{Hypergraphs and relational hyperevents}

In this paper we discuss models for coauthor networks given by (1) a potentially time-varying population of scientists $V_t$, representing the actors, or nodes, of the network who could interact at time $t$, and (2) a sequence of publication events $E=(e_1,\dots,e_N)$, each representing a published scientific paper, coauthored by actors from the given population. The underlying data structure of RHEM and RHOM are \emph{hypergraphs} \citep{berge1989hypergraphs,wasserman1994social,bretto2013hypergraph} which are a generalization of graphs to polyadic, or multi-actor, interaction. Formally, a hypergraph $G=(V,H)$ is given by a set of nodes $V$ and a set of \emph{hyperedges} $H\subseteq\mathcal{P}(V)$. Each hyperedge $h\in H$ is a subset of nodes $h\subseteq V$ of any size. (If all hyperedges have size two, the hypergraph is a graph containing edges that relate exactly two nodes.)

A \emph{relational hyperevent} \citep{lerner2019rem,lerner2021dynamic} $e\in E$, representing a publication event in a coauthor network (i.\,e., a published paper), is given by a tuple
\[
e=(t_e,h_e,x_e,y_e)\enspace,
\]
where $t_e$ is the time of the event (i.\,e., the publication date), $h_e\subseteq V_{t_e}$ is the hyperedge on which the event occurs, representing the authors of the paper, $x_e$ is the event type and/or event weight (see the explanation given below), and $y_e$ is the \emph{relational outcome}, representing the impact of the paper. The crucial difference between the event type (or weight) $x_e$ and the relational outcome $y_e$ is that $x_e$ characterizes the event (that is, the paper), while $y_e$ results from the event. Examples for the event type or weight include the scientific discipline or the journal in which the paper has been published, the impact factor of the journal, or the topic of the paper, for instance, given by its keywords.\footnote{For instance, in the empirical data that we analyze in this paper (see Sect.~\ref{sec:empirical}) the event type $x_e$ is a categorical (multinomial) variable giving the authors' scientific discipline, which is one of \emph{Physics}, \emph{Medicine}, or \emph{Social Science}.} These variables can have an impact on the probability that an event of the given type occurs on a given hyperedge, since groups of scientists ``belong to'' certain scientific disciplines, some scientists are more likely to publish in high-impact journals, scientists have preferred topics on which they publish, and so on. The relational outcome $y_e$, on the other hand, represents the impact of a published paper,\footnote{For instance, in our empirical data, the relational outcome $y_e$ is the normalized number of citations that the paper has gathered by the time of data collection, where the normalization controls for the scientific discipline and the publication date, see Sect.~\ref{sec:rom} for details.} is unknown at the event time (publication date) but rather results from the event. The relational outcome is a measure of performance of the team of authors and represents the response variable in RHOM (see Sect.~\ref{sec:rom}) -- models that can specify the expected impact of papers dependent on previous events in the coauthor network.

\begin{figure}
  \begin{center}
    \begin{tikzpicture}
      \node at (-4,0) (twomode) {\footnotesize
        \begin{tikzpicture}[scale=0.9]
          \tikzstyle{act2}=[circle,minimum size=1mm];
          \tikzstyle{evt}=[rectangle,minimum size=2.5mm];
          \node[act2] at (0,4) (a) {$A$};
          \node[act2] at (0,3.4) (b) {$B$};
          \node[act2] at (0,2.8) (c) {$C$};
          \node[act2] at (0,2.2) (d) {$D$};
          \node[act2] at (0,1.6) (e) {$E$};
          \node[act2] at (0,1.0) (f) {$F$};
          \node[act2] at (0,0.4) (g) {$G$};
          \node[act2] at (0,-0.2) (h) {$H$};
          \node[act2] at (0,-0.8) (i) {$I$};

          \node[evt,fill=grey1] at (3,3.9) (e1) {$e_1$};
          \node[evt,fill=grey2] at (3,2.8) (e2) {$e_2$};
          \node[evt,fill=grey3] at (3,1.7) (e3) {$e_3$};
          \node[evt,fill=grey4] at (3,0.6) (e4) {$e_4$};
          \node[evt,fill=grey5] at (3,-0.5) (e5) {$e_5$};

          \draw[color=grey,line width=1pt] (a) to (e1);
          \draw[color=grey,line width=1pt] (b) to (e1);
          \draw[color=grey,line width=1pt] (a) to (e2);
          \draw[color=grey,line width=1pt] (c) to (e2);
          \draw[color=grey,line width=1pt] (b) to (e3);
          \draw[color=grey,line width=1pt] (c) to (e3);
          \draw[color=grey,line width=1pt] (c) to (e4);
          \draw[color=grey,line width=1pt] (d) to (e4);
          \draw[color=grey,line width=1pt] (e) to (e4);
          \draw[color=grey,line width=1pt] (f) to (e4);
          \draw[color=grey,line width=1pt] (g) to (e4);
          \draw[color=grey,line width=1pt] (f) to (e5);
          \draw[color=grey,line width=1pt] (h) to (e5);
          \draw[color=grey,line width=1pt] (i) to (e5);
      \end{tikzpicture}};
      \node at (4,0) (hyper) {\footnotesize
        \begin{tikzpicture}[scale=0.7]
          \tikzstyle{actor}=[circle,minimum size=1mm];
          \node[actor] at (-4,-1.5) (a) {$A$};
          \node[actor] at (-4,1.5) (b) {$B$};
          \node[actor] at (-2,0) (c) {$C$};
          \node[actor] at (-1,-2) (d) {$D$};
          \node[actor] at (1,-2) (e) {$E$};
          \node[actor] at (2,0) (f) {$F$};
          \node[actor] at (0,1.4) (g) {$G$};
          \node[actor] at (4,-1.5) (h) {$H$};
          \node[actor] at (4,1.5) (i) {$I$};

          \draw[color=darkgrey,line width=1pt,dashed] (a) to (b);
          \draw[color=darkgrey,line width=1pt,dashed] (a) to (c);
          \draw[color=darkgrey,line width=1pt,dashed] (b) to (c);

          \draw[color=darkgrey,line width=1pt,dashed] (f) to (i);
          \draw[color=darkgrey,line width=1pt,dashed] (f) to (h);
          \draw[color=darkgrey,line width=1pt,dashed] (h) to (i);

          \draw[color=darkgrey,line width=1pt,dashed] (c) to (d);
          \draw[color=darkgrey,line width=1pt,dashed] (c) to (e);
          \draw[color=darkgrey,line width=1pt,dashed] (c) to (f);
          \draw[color=darkgrey,line width=1pt,dashed] (c) to (g);
          \draw[color=darkgrey,line width=1pt,dashed] (d) to (e);
          \draw[color=darkgrey,line width=1pt,dashed] (d) to (f);
          \draw[color=darkgrey,line width=1pt,dashed] (d) to (g);
          \draw[color=darkgrey,line width=1pt,dashed] (e) to (f);
          \draw[color=darkgrey,line width=1pt,dashed] (e) to (g);
          \draw[color=darkgrey,line width=1pt,dashed] (f) to (g);

          \begin{pgfonlayer}{background}
            \foreach \nodename in {a,b,c,d,e,f,g,h,i} {
              \coordinate (\nodename') at (\nodename);
            }
            \path[fill=grey1,draw=grey1,line width=0.55cm, line cap=round, line join=round] 
            (a') to (b') to (a') -- cycle;
            \path[fill=grey2,draw=grey2,line width=0.5cm, line cap=round, line join=round] 
            (a') to (c') to (a') -- cycle;
            \path[fill=grey3,draw=grey3,line width=0.45cm, line cap=round, line join=round] 
            (b') to (c') to (b') -- cycle;
            \path[fill=grey4,draw=grey4,line width=0.7cm, line cap=round, line join=round] 
            (c') to (d') to (e') to (f') to (g') to (c') -- cycle;
            \path[fill=grey5,draw=grey5,line width=0.7cm, line cap=round, line join=round] 
            (f') to (h') to (i') to (f') -- cycle;
          \end{pgfonlayer}
        \end{tikzpicture}
      };
    \end{tikzpicture}
  \end{center}
  \[
  e_1=(t_1,\{A,B\});\;e_2=(t_2,\{A,C\});\;e_3=(t_3,\{B,C\});\;e_4=(t_4,\{C,D,E,F,G\});\;e_5=(t_5,\{F,H,I\})
  \]
    \caption{\emph{Bottom:} a list of five hyperevents $e_i=(t_i,h_i)$ representing publication events at event times $t_1<\dots,t_5$. Authors are denoted by letters $A,\dots,I$; event type and relational outcome are not given in this example. \emph{Left:} representation of the hyperevents as a two-mode ``author-paper'' network. Papers (that is, events) are displayed as rectangular nodes labeled $e_1,\dots,e_5$ and are connected to their authors by solid lines. The event nodes are ordered from top to bottom by publication time and older papers are represented by nodes with a darker shade. \emph{Right:} representation of the hyperevents as a hypergraph. Hyperedges represent papers (that is, events) and are displayed as gray-shaded convex hulls enclosing their authors (gray shades of hyperedges match those of the event nodes in the two-mode network). Dashed lines represent author-author ties in the one-mode projection.}
    \label{fig:hyperevents}
\end{figure}

There is a canonical one-to-one mapping between hypergraphs and two-mode networks \citep{seidman1981structures}; also see the illustration in Fig.~\ref{fig:hyperevents}.
Concretely, a hypergraph $(V,H)$ is associated with a two-mode network containing two node sets: the ``actor nodes'' $V$, which are identical with the nodes of the hypergraph, and the ``event nodes'' (or ``paper nodes'') $H$, which are identical with the hyperedges of the hypergraph.
An actor $v\in V$ is connected to an event node $h\in H$ by an edge in the two-mode network if the hyperedge $h$ contains the actor $v$ in the hypergraph.\footnote{We note that the same set of actors can coauthor more than one paper, that is, can experience more than one event. Thus, our hypergraphs are \emph{multi-hypergraphs} where hyperedges have multiplicities. Multiple copies of the same hyperedge are mapped to structurally equivalent event nodes in the two-mode network, that is, to event nodes that are adjacent to the same set of actor nodes. For sake of simplicity, we left out this aspect when discussing the equivalence of hypergraphs and two-mode networks.}

Two-mode networks, or the equivalent hypergraphs, are appropriate network structures for representing the duality of actors and groups \citep{breiger1974duality} in which actors are characterized by the groups they are members of and, dually, groups are characterized by their members. Such group membership is often reflected in observable attendance of actors in joint events, for instance, social gatherings \citep{davis1941deep,freeman2003finding}, scheduled meetings \citep{lerner2021dynamic}, or joint projects tackled by teams as in this paper. Such event networks have two characteristics motivating their analysis with RHEM. First, events often involve multiple actors at the same time, rather than giving rise to independent dyadic interaction. Second, events (as the name itself suggests) typically happen at given points or intervals in time and past events plausibly shape the distribution of future events. RHEM, as proposed in \citet{lerner2021dynamic}, are models considering both aspects. In the setting of this paper, events have an additional component: their outcome. We consider relational hyperevents representing a team that tackles a given task, provides a service, or produces a product and where these events are associated with a measurable outcome. Besides scientific collaboration, other potential application areas include project teams in companies developing a new product or registering a patent \citep{trajtenberg1990penny}, teams of artists and other staff producing a movie \citep{ravasz2003hierarchical}, sports teams \citep{mukherjee2019prior}, or medical teams performing a given surgery \citep{pallotti2020lost}. Relational outcome can serve both as an explanatory variable for the occurrence of future events (motivating to extent RHEM by novel hyperedge statistics based on the outcome of past events) and as a response variable (motivating the development of novel RHOM that can explain outcome of joint events).

The one-mode projection of a two-mode actor-event network, or the equivalent hypergraph, is a graph whose set of nodes is identical with the actor nodes of the two-mode network and where two actors are connected by an edge, if both are connected to a common event node, that is, if both coauthored at least one paper. There is also a dual one-mode projection that considers events as nodes and connects two events that share common actors; this kind of one-mode projection has been considered, for instance, in studies of organizational communication networks \citep{blaschke2012organizations}. It is well known that one-mode projections do not uniquely represent two-mode networks (or the equivalent hypergraphs); e.\,g., \citep{chodrow2020annotated}. This can be seen, for instance, by the triads $\{A,B,C\}$ and $\{F,H,I\}$ from Fig.~\ref{fig:hyperevents} which are identically connected in the one-mode projection, but structurally different in the two-mode network and in the hypergraph.

\subsection{Insufficiency of common statistical network models coauthor networks}

We first recall that modeling one-mode projections, for instance with (temporal) ERGM or SAOM \citep{lkr-ergmsn-13,krivitsky2014separable,s-mlnd-05} has its drawbacks. As discussed above, one-mode projections do not uniquely represent two-mode networks so that, for instance, the difference between the triads $\{A,B,C\}$ and $\{F,H,I\}$ from Fig.~\ref{fig:hyperevents} could get lost. Moreover, one-mode projections create an abundance of closed triangles, especially when papers have many coauthors. For instance, the single event $e_4=(t_4,\{C,D,E,F,G\})$ in Fig.~\ref{fig:hyperevents} yields ${5 \choose 3}=10$ closed triangles. A paper with one hundred authors (the largest that we have in our empirical data considered in this paper) creates an excessive number of ${100 \choose 3}=161,700$ closed triangles. Thus, modeling one-mode projections fails to clarify whether an over-representation of closed triangles can be attributed to triadic closure or to the publication of multi-author papers. In the example, from Fig.~\ref{fig:hyperevents}, the sequence of events $e_1=(t_1,\{A,B\});\;e_2=(t_2,\{A,C\});\;e_3=(t_3,\{B,C\})$ points to triadic closure since the third event $e_3=(t_3,\{B,C\})$ closes a two-path from $B$ over $A$ to $C$. In contrast, the event $e_5=(t_5,\{F,H,I\})$ -- which results in a structurally equivalent configuration in the one-mode projection as the three events $e_1,e_2,e_3$ -- cannot be attributed to triadic closure since it did not close any two-path that was open at the time of the event.

We therefore recommend to model coauthor networks as two-mode networks -- or the equivalent hypergraphs. For the network model families ERGM and SAOM variants for two-mode networks have been defined \citep{wang2013exponential,koskinen2012modelling,snijders2013model}. However, these model frameworks typically take the node set as given and model which node pairs are connected by social ties. This could be problematic for two-mode coauthor networks where the ``event nodes'' (or paper nodes) are not exogenously given but result endogenously from social interaction. Indeed, scientific papers are not created in isolation and scientists cannot later decide to connect to, or disconnect from, existing paper nodes. It is rather the case that scientific papers are endogenously created by the interaction among scientists.

For these reasons we consider it preferable to treat scientific papers not as nodes (of a two-mode network) but rather as relational events that result from interaction among groups of scientists. However, REM \citep{b-refsa-08,bls-ness-09,lbsb-mftien-13,perry2013point,vpr-remslm-15,sb-iat-17,lerner2020reliability} typically specify \emph{dyadic} event rates, associated with pairs of nodes. To cope with the polyadic interaction in coauthor networks we recall RHEM \citep{lerner2021dynamic} which are a generalization of REM to multi-actor events and develop them further to take into account relational outcome variables associated with hyperevents in coauthor networks. Modeling coauthor networks with RHEM is preferable to applying dyadic REM since the latter could not account for higher-order dependencies in joint publication events. Modeling pairwise interaction in the one-mode projection would suffer from structural artifacts, such as the abundance of closed triangles, described above. Modeling dyadic paper-author events in the two-mode network with REM would assume independence of these dyadic events, which would be invalid in general. Moreover, dyadic event rates specified for individual paper-author pairs could just depend on characteristics of single authors (perhaps in combination with the paper node) -- but could not be functions of pairs or larger sets of coauthors. This would preclude, for instance, the analysis of familiarity, prior shared success, or closure effects. A concrete example of an effect in RHEM that cannot be specified with REM -- neither in the one-mode nor in the two-mode representation -- is subset repetition (or prior shared success) of order three or higher, formally defined in Sect.~\ref{sec:statistics}. Albeit going beyond dyadic relational events, RHEM for meeting events as they have been proposed in \citet{lerner2021dynamic} would be incomplete models for coauthor networks for two reasons. First, these previous RHEM do not specify event rates dependent on the relational outcome of past hyperevents (note that in the example data from \citet{lerner2021dynamic} there is not outcome associated with meeting events). Secondly, RHEM, as proposed in \citet{lerner2021dynamic} explain the occurrence of hyperevents but not their outcome which is a limitation when analyzing coauthor networks -- a limitation that will be overcome by the newly proposed RHOM in this paper.

\section{Network models for scientific collaboration and impact}
\label{sec:model}

In this section we elaborate the RHEM framework proposed in \citet{lerner2021dynamic} for coauthor networks, taking into account that relational outcome variables associated with publication events can influence the probability of future collaboration, and propose RHOM to explain scientific impact of published papers. RHEM and RHOM provide models for sequences of publication events $E=(e_1,\dots,e_N)$, where each event $e\in E$ is a tuple
\[
e=(t_e,h_e,x_e,y_e)\enspace,
\]
comprising publication time $t_e$, set of authors $h_e$, event type $x_e$, and relational outcome $y_e$. RHEM for event intensities \citep{lerner2021dynamic}, recalled in Sect.~\ref{sec:rhem}, explain publication rates associated with hyperedges (i.\,e., associated with groups of scientists of any size) and RHOM (specified in Sect.~\ref{sec:rom}) explain the expected impact of published papers. In both models, response variables are stochastic functions of previous events on the same or incident hyperedges.

\subsection{RHEM: modeling scientific collaboration intensity}
\label{sec:rhem}

Generalizing definitions of dyadic REM \citep{perry2013point}, \citet{lerner2021dynamic} specify RHEM as marked point processes on hyperedges. In contrast to dyadic REM, point processes in RHEM are labeled with hyperedges comprising any number of nodes, rather than labeled with dyads comprising exactly two nodes. Thus, for the given sequence of publication events $E=(e_1,\dots,e_N)$, RHEM specify the \emph{intensity} (also denoted as \emph{event rate} or \emph{publication rate} in this paper) $\lambda(t_e,h)$ for each hyperedge $h\in R_{t_e}$ in the given \emph{risk sets} $R_{t_e}\subseteq\mathcal{P}(V_{t_e})$ at the event times $t_e$. Intuitively, the intensity, or publication rate, $\lambda(t_e,h)$ is the expected number of papers co-published by $h$ in a time interval of unit length starting at $t_e$ \citep{lerner2021dynamic}. The risk set $R_{t_e}$ contains those hyperedges $h\subseteq V_{t_e}$ that could potentially publish a joint paper at $t_e$ and that we want to compare with the hyperedge $h_e$ of the observed event $e$. More specifically, the risk sets $R_{t_e}$ that we consider in RHEM in this paper contain a random sample of hyperedges of the same size as the hyperedge of the observed event $e$; see details below.

RHEM specify these publication rates, among others, dependent on previous events on $h$ or incident hyperedges. Following \citet{lerner2021dynamic}, the relative event rate $\lambda_1$ is specified within the framework of the Cox proportional hazard model \citep{cox1972regression} as a function of \emph{hyperedge statistics} $s(t,h,G[E;t])=[s_1(t,h,G[E;t]),\dots,s_k(t,h,G[E;t])]\in\mathbb{R}^k$, quantifying various aspects of how the hyperedge $h$ is embedded into the \emph{network of past events} $G[E;t]$ \citep{bls-ness-09} at time $t$, and a vector of associated parameters $\theta=(\theta_1,\dots,\theta_k)\in\mathbb{R}^k$, describing which of these statistics increase or decrease the relative event rate:
\[
\lambda_1(t,h,\theta,G[E;t])=\exp\left(\sum_{i=1}^k\theta_i\cdot s_i(t,h,G[E;t])\right)\enspace.
\]
Specific statistics that we include in the vector $s(t,h,G[E;t])$ in the empirical analysis of this paper are introduced in Sect.~\ref{sec:statistics}. In this paper we assume that $G[E;t]$ captures all relevant information that shapes the probability distribution for publication events at time $t$. This information can include previous papers, published at $t'<t$, but also endogenously given covariates of actors or hyperedges, such as demographic variables or institutional affiliation. 


Parameters $\theta$ in the Cox proportional hazard model are estimated to maximize the partial likelihood $L$ based on the observed sequence of publication events $E$:
\begin{equation}
  \label{eq:likelihood}
  L(\theta)=\prod_{e\in E}\frac{\lambda_1(t_{e},h_{e},\theta,G[E;t_{e}])}
  {\sum_{h\in R_{t_{e}}}\lambda_1(t_{e},h,\theta,G[E;t_{e}])}\enspace.
\end{equation}
The likelihood given in Eq.~(\ref{eq:likelihood}) does not explicitly mention the event type $x_e$. In our empirical analysis the event type is a categorical variable, giving the scientific discipline of the paper (\emph{Physics}, \emph{Medicine}, or \emph{Social Science}). In this paper, we assume for simplicity that the three disciplines represent separate independent networks, so that a publication event in, say, physics does not depend on previous publication events in medicine or social science. We then fit two types of models: the first assumes that publications in the three disciplines are drawn from the same model (having the same parameters). This means that each of the three disciplines yields a likelihood function as given in Eq.~(\ref{eq:likelihood}) and we estimate one vector of parameters to maximize the joint likelihood which is the product of the three discipline-specific likelihoods. In the second type of models we assume that publications in the three disciplines are drawn from separate models (that can have different parameters). Thus, each of the three disciplines yields a likelihood function of the form given in Eq.~(\ref{eq:likelihood}) and we estimate for each discipline a separate parameter vector by individually maximizing this likelihood. In general, if we did not split the event sequence by the event type, the relative event rate $\lambda_1$ and the hyperedge statistics $s_i$ would contain the event type $x_e$ as an additional argument. 

The relative event rate $\lambda_1$ does not depend on the relational outcome $y_e$. Indeed, as we discussed earlier, the relational outcome of event $e$ has no influence on the occurrence of events at time $t_e$ (since it is unknown at the event time). However, $y_e$ can have an influence on the occurrence of future events at $t>t_e$. Moreover, the relational outcome $y_e$ is treated as the result of an event and is modeled by RHOM as explained in Sect.~\ref{sec:rom}.

Since the size of the full risk set is exponential in the number of actors, we apply case-control sampling \citep{bgl-mascdcphm-95} to ensure computational tractability. In case-control sampling we sample for each observed event (``case'') $e=(t_e,h_e)$ a constant number of hyperedges from the risk set $R_{t_e}$ that could have experienced an event at $t_e$ but did not (``controls'' or ``non-events''). Sampling of non-event hyperedges from the risk set is done uniformly at random. Case-control sampling has been applied to dyadic REM \citep{vpr-remslm-15,lerner2020reliability} and to RHEM \citep{lerner2019rem,lerner2021dynamic}. Parameters are estimated from sampled likelihood functions that have the same form as that given in Eq.~(\ref{eq:likelihood}) but where the risk sets contain only the sampled controls and the hyperedges of observed events. Estimating parameters of Cox proportional hazard models via case-control sampling is a consistent estimator \citep{bgl-mascdcphm-95}. The experiments reported in \citet{lerner2020reliability} revealed that REM parameters on large event networks can be reliably estimated with sampled observations containing some tens of thousands of events and some hundreds of thousands of controls -- even if the distribution of statistics over the observations is very skewed. For more well-behaved explanatory variables, sample sizes could even be much smaller. Since we fit models on data with about ten times as many observations, we believe that the error introduced by sampling does not distort our findings qualitatively. We also follow the recommendation of \citet{lerner2020reliability} and reestimate RHEM parameters on ten independent samples of the same size. The standard deviation of parameters over samples was of about the same size as the standard errors, which yields further confidence in the reliability of empirical findings.

We further restrict the risk set $R_{t_e}$ associated with the event $e=(t_e,h_e)$ to contain only hyperedges with the same number of actors as the hyperedge $h_e$ of the observed event. Thus, we estimate \emph{conditional-size RHEM}, which has been advocated in \citet{lerner2019rem,lerner2021dynamic}, since baseline event intensities depend to a high degree on the hyperedge size (note that there are by several orders of magnitude more hyperedges of size, say, ten than of size two). However, in our illustrating application (see Sect.~\ref{sec:empirical}) the argument for conditioning on hyperedge size is even simpler. Drawing controls from the unconstrained risk set would give us hyperedges that contain in expectation up to almost 200,000 actors -- which would be an absurd number of coauthors. There would be no point in comparing explanatory variables (hyperedge statistics) associated with hyperedges of observed events, which contain typically between one and 20 actors, with the explanatory variables associated with hyperedges of such absurdly large size. Conditioning on hyperedge size is a way to ensure that observed events are matched with alternative (non-event) hyperedges with which they are better comparable.

However, it has to be kept in mind that conditioning on hyperedge size actually changes the likelihood function -- and, thus, the maximum likelihood estimates of the parameters -- and has implications for the interpretation of findings \citep{lerner2021dynamic}. For instance, a finding such as ``prior shared success tends to increase collaboration frequencies'' does not imply a tendency to add more and more authors with whom the current authors share prior success (since just adding authors would change the hyperedge size) -- it rather implies a tendency to drop an author with whom the others have little prior success in exchange for including an author with whom the others have higher prior shared success (keeping hyperedge size constant). We further note that an alternative to condition on hyperedge size would be to control for it in a way that the expected size (according to the model) is comparable to the observed one -- rather than constrained to be equal. We come back to this aspect when discussing future work at the end of this paper.

\subsection{RHOM: modeling the impact of papers}
\label{sec:rom}

Relational hyperevent outcome models (RHOM) can explain the impact, or relational outcome, $y_e$ of published papers, represented as publication events $e=(t_e,h_e,x_e,y_e)$. RHOM go beyond RHEM \citep{lerner2021dynamic} but are related with the latter in the sense that both are models for response variables associated with hyperedges. The difference is that RHEM specify the relative rate $\lambda_1(t,h)$ for all hyperedges $h$ in the risk set, while RHOM specify conditional probability distributions $f(y_e\,|\,t_e,h_e)$ for the relational outcome $y_e$, given that there is an event on the hyperedge $h_e$ at time $t_e$; compare \citet{lerner2019rem}.

In our empirical study, relational outcome is the impact of a published paper quantified by the normalized number of citations that the paper has received at the time of data collection. The normalization considers the discipline of the paper and the year of publication. Specifically, if $c_e$ denotes the observed number of citations of the paper represented by $e$ at the time of data collection, we subtract from this raw citation count the average number of citations taken over all papers from the same discipline and the same year of publication as $e$. That is, the impact $y_e$ of the paper represented by publication event $e$ is defined as
\begin{equation}\label{eq:impact}
y_e=c_e-\frac{\sum_{e'\colon x_{e'}=x_e\wedge t_{e'}=t_e}c_{e'}}{|\{e'\in E\colon x_{e'}=x_e\wedge t_{e'}=t_e\}|}\enspace.
\end{equation}
The impact $y_e$ is positive (negative) if paper $e$ gathers more (fewer) citations than the average paper from the same discipline published in the same year. 

RHOM \citep{lerner2019rem} specify the likelihood of an observed sequence of relational hyperevents $E$ by
\begin{equation}
  \label{eq:likelihood_rhom}
  L(\theta)=\prod_{e\in E}f(y_e\,|\,t_{e},h_{e},\theta,G[E;t_{e}])\enspace,
\end{equation}
where $f$ is a suitably chosen distribution for the relational outcome $y_e$, typically from the family of generalized linear models (GLM). Despite this simplicity, RHOM can control for certain types of non-independence among observations in a way that goes beyond what is possible in typical GLM. For instance, the impact of a paper is likely to be dependent on the impact of earlier papers published by the same or overlapping authors. RHOM can control for this kind of dependence by specifying the conditional distribution $f(y_e\,|\,t_{e},h_{e},\theta,G[E;t_{e}])$ as a function of previous publication events where the precise structure of dependence is shaped by the inclusion of hyperedge statistics defined in Sect.~\ref{sec:statistics}.

We model the impact of papers by linear regression (ordinary least squares), where explanatory variables for the relational outcome $y_e$ can be a function of previous events (that is, papers published before the year $t_e$) in the same network. Thus, RHOM can model relational outcome dependent on previous publication events on the same or incident hyperedges in a similar way as RHEM can model event rates dependent on previous events. As explanatory variables we typically use the same hyperedge statistics as in the model for event rates. Indeed, specifying event rates (explaining which group of scientists co-publishes papers) and relational outcome (impact of published papers) by the same explanatory variables allows to assess consistency or effectivity of mechanisms explaining scientific team formation. The overarching question is whether factors that increase (or decrease) the likelihood of co-publication also tend to increase (or decrease) the expected impact of the published papers. If this holds true, then it would be a sign that scientists have a tendency to assemble into successful teams. In contrast, if some explanatory variable had the opposite effect on team formation and performance, it would point to adverse selection in team formation in the sense that scientists would have a tendency to form unsuccessful teams -- or a reluctance to form potentially successful teams.

\subsection{Hyperedge statistics for the specification of network effects in RHEM and RHOM}
\label{sec:statistics}

Hyperedge statistics operationalize hypothetical effects in models explaining publication rates (Sect.~\ref{sec:rhem}), or impact of published papers (Sect.~\ref{sec:rom}), respectively. In this paper we use two types of network effects modeling (1) dependence on the \emph{occurrence} of past publication events, operationalized by hyperedge statistics already defined in \citet{lerner2021dynamic}, and (2) dependence on the \emph{impact} (or relational outcome) of past publications. For the given sequence of events $E$ and a point in time $t$, we denote by $E_{<t}=\{e\in E\colon\; t_e<t\}$ the past events, that is, those events that happen strictly before $t$.

\paragraph{Subset repetition.}
The first, rather obvious, family of network effects accounts for repeated (co-)authorship. More specifically, if a hyperedge $h\subseteq V$ (that is, a set of scientists) has already published a joint paper, potentially together with other scientists outside of $h$, then we expect that $h$ has an increased probability to co-publish again in the future, potentially together with yet other scientists. The interpretation of this effect depends on the size of the repeated hyperedge $h$ (dubbed the \emph{order} of the subset repetition effect). Subset repetition of order one accounts for the hypothetical effect that scientists who published larger numbers of papers in the past, will publish more in the future. (Both the past and the future papers might be coauthored with others.) Such an effect -- if empirically supported -- would point to preferential attachment \citep{kronegger2012collaboration} in which scientists who were more productive in the past, accumulate publications at a higher rate. Subset repetition of order two or more accounts for \emph{familiarity} effects of different order. Subset repetition of order two tests the hypothetical effect that \emph{pairs} of scientists are more likely to coauthor in the future, if they have coauthored in the past. Subset repetition of order three hypothesizes that \emph{triads} of scientists who have jointly published before are more likely to co-publish again.

Subset repetition effects can be illustrated with the two triads $\{A,B,C\}$ and $\{F,H,I\}$ from the example given in Fig.~\ref{fig:hyperevents}. We observe that each pair of authors from $\{A,B,C\}$ has coauthored one paper and the same holds true for each pair of authors from $\{F,H,I\}$. Thus, a model accounting only for subset repetition of order two would assess these two triads identically. However, the second triad has published a joint paper (written by all three members), but the first triad did not. Thus, a model accounting for subset repetition of order three could assess these two triads differently. This example illustrates a hypothetical network effect in coauthor networks that could not be accounted for by dyadic REM -- neither in the one-mode representation, nor in the two-mode representation.

Formally, subset repetition is defined in two steps \citep{lerner2021dynamic}. First the \emph{hyperedge degree} $deg(t;h';G[E;t])$ counts how many papers have been coauthored (potentially together with yet other scientists outside of $h'$) by all members of $h'$ before time $t$:
\[
deg(t,h',G[E;t])=\sum_{e\in E_{<t}}\chi(h'\subseteq h_e)\enspace.
\]
(The characteristic function $\chi$ is one if the argument is true and zero else.) For a given integer $p\in\mathbb{N}$ (specifying the size of subsets that are to be repeated), subset repetition of order $p$ is defined by
\[
sub.rep^{(p)}(t,h,G[E;t])=\sum_{h'\in{h \choose p}}deg(t,h',G[E;t])\cdot\frac{1}{{|h|\choose p}}\enspace,
\]
where we write ${h \choose p}=\{h'\subseteq h\colon\;|h'|=p\}$ to denote all subsets of $h$ of size $p$. In words, subset repetition of order $p$ is the average hyperedge degree over all subsets of size $p$ of the focal hyperedge $h$. In our empirical analysis, we fit models including subset repetition of order one, two, and three; in the appendix we fit additional models with subset repetition up to order ten. We note that models specified with subset repetition up to a maximal order $p$ (e.\,g., $p=3$) are still applicable to larger publication events, i.\,e., papers exceeding $p$ authors. This is because subset repetition allows that given subsets of order $p$ repeat co-authoring events \emph{potentially together with yet other authors}. 

\paragraph{Closure.}
Triadic closure effects predict that scientists are more likely to coauthor if they have previously coauthored with common third actors \citep{ferligoj2015scientific}. Triadic closure leads to an over-representation of closed triangles in one-mode projections of coauthor networks. However, as discussed above, the reverse implication does not hold: an over-representation of closed triangles in one-mode projections of coauthor networks can also result from papers with many authors and/or from the tendency to partially repeat such multi-actor collaborations.

Quantitatively, the statistic $closure(t,h,G[E;t])$ \citep{lerner2021dynamic}, defined below, iterates over all unordered pairs $\{u,v\}\subseteq h$ and for each of these pairs, it iterates over all scientists $w$ (within or outside of $h$) that are different from $u$ and from $v$. We then take the minimum number of previous joint publications of $\{u,w\}$ and $\{v,w\}$ as a measure for how strongly $u$ and $v$ are indirectly connected via the third actor $w$. Summing over all third actors $w$ yields a measure for how strongly $u$ and $v$ are indirectly connected and we average this measure over all unordered pairs $\{u,v\}\subseteq h$ to quantify how much a joint publication event on the hyperedge $h$ would close indirect collaborations. In formulas, the closure statistic is defined by
\[
closure(t,h,G[E;t])=\sum_{\{u,v\}\in{h \choose 2}\wedge w\not=u,v}
\min[deg(t,\{u,w\}),deg(t,\{v,w\})]/{|h|\choose 2}\enspace,
\]
where we drop the argument $G[E;t]$ in the hyperedge degree for brevity. We note that, similar to subset repetition of order $p$, closure is a hyperedge statistic for hyperedges of any size, in particular, closure is not restricted to hyperedges of size three.

\begin{figure}
  \begin{center}
    \begin{tikzpicture}
      \node at (-3.5,0) (hyper) {\footnotesize
        \begin{tikzpicture}[scale=0.6]
          \tikzstyle{actor}=[circle,minimum size=1mm];
          \node[actor] at (-4,-1.5) (a) {$A$};
          \node[actor] at (-4,1.5) (b) {$B$};
          \node[actor] at (-2,0) (c) {$C$};
          \node[actor] at (-1,-2) (d) {$D$};
          \node[actor] at (1,-2) (e) {$E$};
          \node[actor] at (2,0) (f) {$F$};
          \node[actor] at (0,1.4) (g) {$G$};
          \node[actor] at (4,-1.5) (h) {$H$};
          \node[actor] at (4,1.5) (i) {$I$};

          \draw[color=darkgrey,line width=1pt,dashed] (a) to (b);
          \draw[color=darkgrey,line width=1pt,dashed] (a) to (c);
          \draw[color=darkgrey,line width=1pt,dashed] (b) to (c);

          \draw[color=darkgrey,line width=1pt,dashed] (f) to (i);
          \draw[color=darkgrey,line width=1pt,dashed] (f) to (h);
          \draw[color=darkgrey,line width=1pt,dashed] (h) to (i);

          \draw[color=darkgrey,line width=1pt,dashed] (c) to (d);
          \draw[color=darkgrey,line width=1pt,dashed] (c) to (e);
          \draw[color=darkgrey,line width=1pt,dashed] (c) to (f);
          \draw[color=darkgrey,line width=1pt,dashed] (c) to (g);
          \draw[color=darkgrey,line width=1pt,dashed] (d) to (e);
          \draw[color=darkgrey,line width=1pt,dashed] (d) to (f);
          \draw[color=darkgrey,line width=1pt,dashed] (d) to (g);
          \draw[color=darkgrey,line width=1pt,dashed] (e) to (f);
          \draw[color=darkgrey,line width=1pt,dashed] (e) to (g);
          \draw[color=darkgrey,line width=1pt,dashed] (f) to (g);

          \begin{pgfonlayer}{background}
            \foreach \nodename in {a,b,c,d,e,f,g,h,i} {
              \coordinate (\nodename') at (\nodename);
            }
            \path[fill=grey1,draw=grey1,line width=0.55cm, line cap=round, line join=round] 
            (a') to (b') to (a') -- cycle;
            \path[fill=grey2,draw=grey2,line width=0.5cm, line cap=round, line join=round] 
            (a') to (c') to (a') -- cycle;
            \path[fill=grey3,draw=grey3,line width=0.45cm, line cap=round, line join=round] 
            (b') to (c') to (b') -- cycle;
            \path[fill=grey4,draw=grey4,line width=0.7cm, line cap=round, line join=round] 
            (c') to (d') to (e') to (f') to (g') to (c') -- cycle;
            \path[fill=grey5,draw=grey5,line width=0.7cm, line cap=round, line join=round] 
            (f') to (h') to (i') to (f') -- cycle;
          \end{pgfonlayer}
        \end{tikzpicture}
      };
      \node at (3.5,0) (hyper2) {\footnotesize
        \begin{tikzpicture}[scale=0.6]
          \tikzstyle{actor}=[circle,minimum size=1mm];
          \node[actor] at (-4,-1.5) (a) {$A$};
          \node[actor] at (-4,1.5) (b) {$B$};
          \node[actor] at (-2,0) (c) {$C$};
          \node[actor] at (-1,-2) (d) {$D$};
          \node[actor] at (1,-2) (e) {$E$};
          \node[actor] at (2,0) (f) {$F$};
          \node[actor] at (0,1.4) (g) {$G$};
          \node[actor] at (4,-1.5) (h) {$H$};
          \node[actor] at (4,1.5) (i) {$I$};

          \draw[color=darkgrey,line width=1pt,dashed] (a) to (b);
          \draw[color=darkgrey,line width=1pt,dashed] (a) to (c);
          \draw[color=darkgrey,line width=1pt,dashed] (b) to (c);

          \draw[color=darkgrey,line width=1pt,dashed] (f) to (i);
          \draw[color=darkgrey,line width=1pt,dashed] (f) to (h);
          \draw[color=darkgrey,line width=1pt,dashed] (h) to (i);

          \draw[color=darkgrey,line width=1pt,dashed] (c) to (d);
          \draw[color=darkgrey,line width=1pt,dashed] (c) to (e);
          \draw[color=darkgrey,line width=1pt,dashed] (c) to (f);
          \draw[color=darkgrey,line width=1pt,dashed] (c) to (g);
          \draw[color=darkgrey,line width=1pt,dashed] (d) to (e);
          \draw[color=darkgrey,line width=1pt,dashed] (d) to (f);
          \draw[color=darkgrey,line width=1pt,dashed] (d) to (g);
          \draw[color=darkgrey,line width=1pt,dashed] (e) to (f);
          \draw[color=darkgrey,line width=1pt,dashed] (e) to (g);
          \draw[color=darkgrey,line width=1pt,dashed] (f) to (g);

          \begin{pgfonlayer}{background}
            \foreach \nodename in {a,b,c,d,e,f,g,h,i} {
              \coordinate (\nodename') at (\nodename);
            }
            \path[fill=grey1,draw=grey1,line width=0.55cm, line cap=round, line join=round] 
            (a') to (b') to (a') -- cycle;
            \path[fill=grey2,draw=grey2,line width=0.5cm, line cap=round, line join=round] 
            (a') to (c') to (a') -- cycle;
            \path[fill=grey3,draw=grey3,line width=0.45cm, line cap=round, line join=round] 
            (b') to (c') to (b') -- cycle;
            \path[fill=grey4,draw=grey4,line width=0.7cm, line cap=round, line join=round] 
            (c') to (d') to (e') to (f') to (g') to (c') -- cycle;
            \path[fill=grey5,draw=grey5,line width=0.7cm, line cap=round, line join=round] 
            (f') to (h') to (i') to (f') -- cycle;

            \path[fill=darkgrey,draw=darkgrey,line width=0.4cm, line cap=round, line join=round] 
            (c') to (f') to (g') to (c') -- cycle;
            \path[fill=darkgrey,draw=darkgrey,line width=0.4cm, line cap=round, line join=round] 
            (d') to (e') to (h') to (d') -- cycle;
            \path[fill=white,draw=white,line width=0.33cm, line cap=round, line join=round] 
            (c') to (f') to (g') to (c') -- cycle;
            \path[fill=white,draw=white,line width=0.33cm, line cap=round, line join=round] 
            (d') to (e') to (h') to (d') -- cycle;
          \end{pgfonlayer}
        \end{tikzpicture}
      };
    \end{tikzpicture}
  \end{center}
  \[
  e_1=(t_1,\{A,B\});\;e_2=(t_2,\{A,C\});\;e_3=(t_3,\{B,C\});\;e_4=(t_4,\{C,D,E,F,G\});\;e_5=(t_5,\{F,H,I\})
  \]
    \caption{Illustrating closure effects. \emph{Bottom:} a list of five hyperevents $e_i=(t_i,h_i)$ representing publication events. \emph{Left:} representation of the hyperevents as a hypergraph. Hyperedges represent papers (that is, events) and are displayed as gray-shaded convex hulls enclosing their authors. Dashed lines represent author-author ties in the one-mode projection (compare Fig.~\ref{fig:hyperevents}). \emph{Right:} representation of the identical hypergraph with two additional hyperedges (possible candidates for future events), $h=\{C,F,G\}$ and $h'=\{D,E,H\}$, represented as white convex hulls with dark borders enclosing their members. An event on $\{D,E,H\}$ would point to a closure effect, while an event on $\{C,F,G\}$ could alternatively be explained by subset repetition.}
    \label{fig:closure}
\end{figure}

Closure -- and its interplay with subset repetition -- is illustrated in Fig.~\ref{fig:closure}. This figure shows how two hyperedges, $h=\{C,F,G\}$ and $h'=\{D,E,H\}$ (which are possible candidates for publication events at a future point in time $t>t_5$), are embedded into the network of past events. Possible future publication events on the two hyperedges $h$ and $h'$ would give varying support for a hypothetical transitive closure effect. An event on $h'=\{D,E,H\}$ would point to closure: all three of its members have co-published before with the common third actor $F$; apart from this, $D$, $E$, and $H$ have relatively few previous events among themselves (only $D$ and $E$ have coauthored one previous paper, the actor $H$ has no collaboration history with the other two). Previous collaboration history is very different for the hyperedge $h=\{C,F,G\}$. While $closure(t,h,G[E;t])$ takes a largely positive value (since members of $h$ have co-published with several common third actors), a possible event on $h$ could alternatively be explained by subset repetition of order two or three (note that all dyads within $\{D,E,H\}$, as well as the whole triad, have coauthored before).

Subset repetition and closure have different  macro-structural implications. Subset repetition of order two or higher leads to the reinforcement of densely connected clusters (like the one formed by $\{C,D,E,F,G\}$) -- and thereby leads to a reinforcement of closed triangles. In contrast, a \emph{positive} closure effect would lead to the formation of new closed triangles, connecting scientists who have not collaborated before. Such an effect would imply that overlapping dense clusters have a tendency to merge. In the example from Fig.~\ref{fig:closure}, the dense clusters $\{C,D,E,F,G\}$ and $\{F,H,I\}$ overlap in the actor $F$; a positive closure effect would make a future publication event on, say, $\{D,E,H\}$ more likely, which would imply that the clusters merge over time. In contrast, in our empirical study we will actually find a \emph{negative} closure effect -- along with positive subset repetition. These two effects together lead to the formation and reinforcement of dense local clusters that might overlap but that have a reluctance to merge. Thus, positive subset repetition together with negative closure provides an explanation for overlapping but stable dense clusters. We will discuss these effects and their implications again in the results section.

\paragraph{Prior individual and shared success.} Future scientific collaboration, as well as future scientific impact, can depend on past individual and shared success, which can be added to RHEM and RHOM via newly defined hyperedge statistics. We distinguish between scientific output and scientific impact. Scientific output refers to the number of papers published by individuals, or co-published by groups, and is captured by the hyperedge degree, defined above. Scientific impact, on the other hand, refers to the (normalized) number of citations gathered by published papers and is quantified in the relational outcome $y_e$ of publication events $e=(t_e,h_e,x_e,y_e)$.

To assess the cumulative prior joint performance of groups of scientists $h'\subseteq V$ of any size, we add up the relational outcome $y_e$ over past publications coauthored by all members of $h'$ (potentially together with other scientists outside of $h'$):
\[
\textit{performance}(t,h',G[E;t])=\sum_{e\in E_{<t}}y_e\cdot\chi(h'\subseteq h_e)\enspace.
\]
Based on this measure of prior joint performance, we define for a positive integer $p$ the \emph{prior success of order} $p$ of a hyperedge $h$ by iterating over all subsets $h'\subseteq h$ of size $p$ and adding up the cumulative prior joint performance of these subsets $h'$. This measure is normalized by the cumulative degree of all those subsets, leading to the statistic\footnote{We resolve $\frac{0}{0}=0$ in this definition, since (groups of) scientists who have never (co-)published before also have no prior (shared) success.} 
\[
prior.success^{(p)}(t,h,G[E;t])=
\frac{\sum_{h'\in{h \choose p}}\textit{performance}(t,h',G[E;t])}
     {\sum_{h'\in{h \choose p}}deg(t,h',G[E;t])}\enspace.
\]
For $p=1$ the statistic $prior.success^{(p)}(t,h,G[E;t])$ gives the average success of individual scientists in $h$ (where it does not matter if this success has been achieved by collaboration with other members of $h$, with scientists outside of $h$, or by solo publications). For $p=2$ the statistic $prior.success^{(p)}(t,h,G[E;t])$ gives the prior \emph{shared} success of pairs of scientists consisting of members of $h$ and for $p=3$ it considers prior shared success of triads within $h$. Similar to subset repetition, it would not be possible to analyze an effect such as prior shared success of order three or higher in a dyadic REM -- or in a relational-outcome variant of a dyadic REM -- since it requires to compute explanatory variables (statistics) associated with hyperedges, rather than with pairs of actors.

\paragraph{Prior success disparity.}
For predicting the success of scientific team work, not only the average prior (shared) success of team members matters, but also its distribution \citep{ahmadpoor2019decoding}. To assess how much members of a scientific team differ with respect to their prior success, we take the standard deviation of the individual performance of team members. In formulas, if we abbreviate $\textit{performance}(t,\{v\},G[E;t])$ by $perf(v)$ and write $\overline{p}(h)$ for the mean performance, $\overline{p}(h)=\sum_{v\in h}perf(v)/|h|$, we define
\[
success.disparity(t,h,G[E;t])=\sqrt{\frac{1}{|h|-1}\sum_{v\in h}[perf(v)-\overline{p}(h)]^2}
\]
where we set the success disparity of hyperedges of size one to zero.

\paragraph{Hyperedge statistics based on covariates.}
In our empirical data used to illustrate RHEM and RHOM for coauthor networks we have no exogenously given covariates of actors or hyperedges -- yet it is straightforward to define hyperedge statistics operationalizing covariate effects in RHEM or RHOM, if such covariates are available. Typical actor-level covariates in a coauthor networks could include gender, age, nationality, geographic location, job position, or tenure; examples of covariates associated with hyperedges (i.\,e., with sets of actors) could be given by shared institutional affiliation or joint membership in research projects. If such covariates are given by numerical variables, they can give rise to hyperedge statistics in exactly the same way as we defined hyperedge statistics based on the hyperedge degree or as a function of the prior joint performance of a hyperedge. With such statistics we could, for instance, assess the effect of mean tenure of the members of a hyperedge, or the effect of tenure disparity. This approach also generalizes to non-numeric (e.\,g., categorical or ordinal) covariates. For instance, given institutional affiliation of actors, we could define a hyperedge statistic equal to the fraction of the members of a hyperedge that have the same institutional affiliation, to assess the effect of institutional homogeneity. Relatedly, \citet{lerner2021dynamic} define hyperedge statistics capturing the first-order effect and the homophily with respect to a binary actor-level covariate.

\paragraph{Decay in the influence of past events over time.} Statistics based on prior events defined so far are cumulative in the sense that they add up the contribution of past events, regardless of how long ago these past events happened. Previous work in REM \citep{lbsb-mftien-13} and RHEM \citep{lerner2021dynamic} suggested to let the influence of past events decay exponentially over time. We do not let the influence of past events decay in the analysis presented in the main text since we assume that past collaboration, or past collaborative success, can have a rather long effect into the future. Since we analyze an observation period of 13 years in the empirical part of this paper, it might well be that events from the very beginning still have an effect at the time of data collection. However, since in other application scenarios a decay over time might be more obvious, or since other studies might analyze longer periods of time, we point out that it is possible to let the effect of past events decay, for instance, as it has been suggested in \citet{lerner2021dynamic}. To check the robustness of our findings we perform an analysis with a decay in the appendix. It turns out that our findings do not change qualitatively.

\section{Illustrative empirical case study}
\label{sec:empirical}

\subsection{Data}
\label{sec:data}

\begin{figure}
  \begin{center}
    \includegraphics[width=0.7\textwidth]{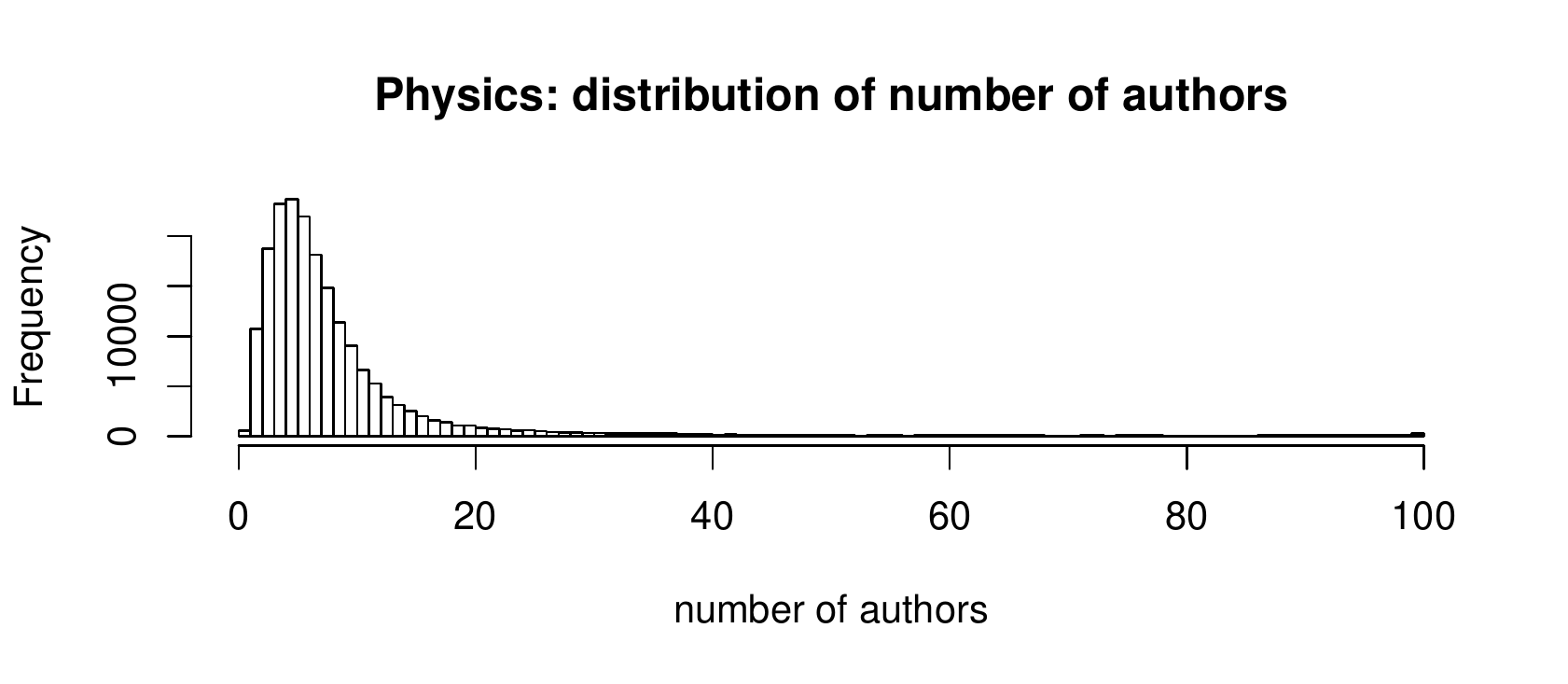}
    \includegraphics[width=0.7\textwidth]{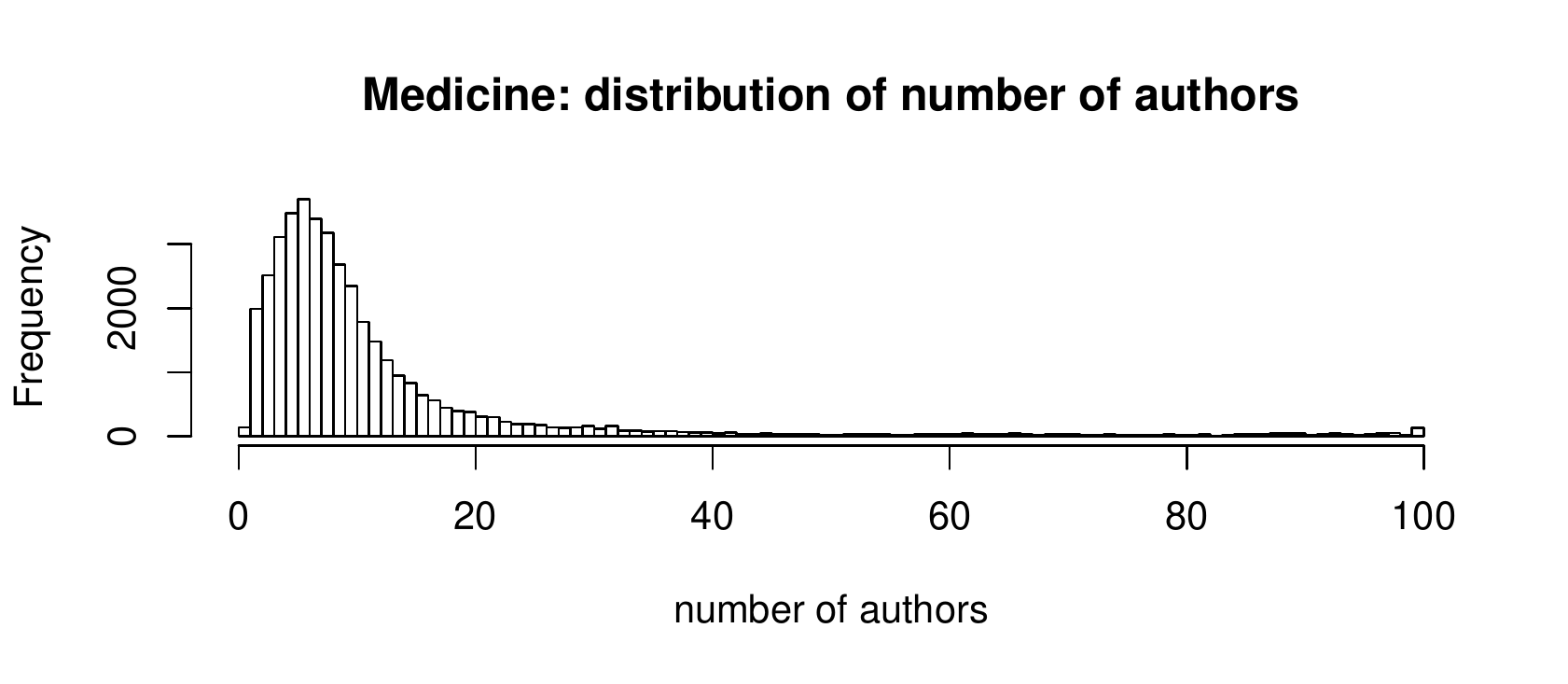}
    \includegraphics[width=0.7\textwidth]{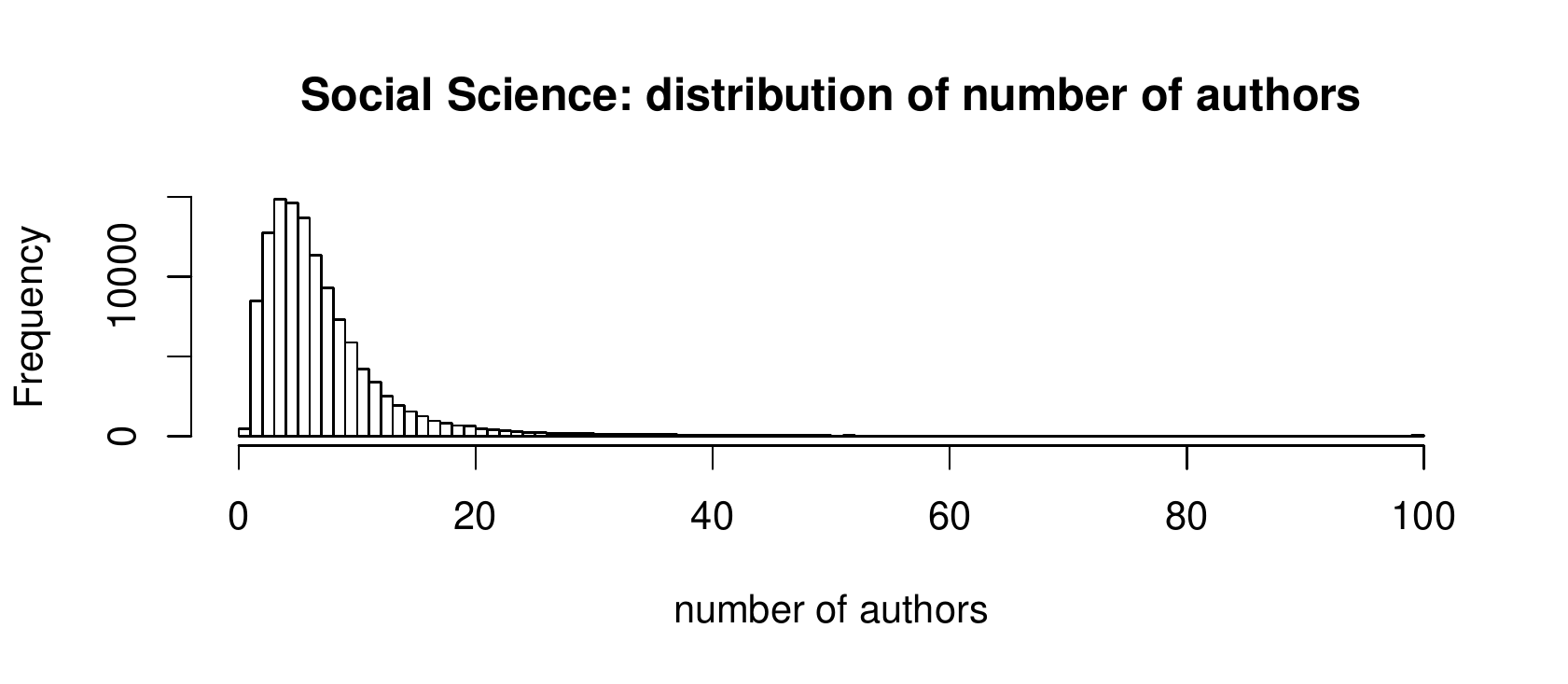}
  \end{center}
  \caption{Histogram of sizes of observed hyperevents in the coauthor data. Hyperevent size is the number of authors of published papers.}
  \label{fig:coauthor_sizes}
\end{figure}

We illustrate the empirical use of our models on a network of publications from the most productive 1,200 EU-based scientists between 2007 and 2019, in each of the following scientific disciplines: physics, medicine, and social science, which yields a total of 3,600 most productive scientists (``seed authors''). These disciplines are considered as three separate networks. A scientist is profiled as working in the field of physics, medicine or social sciences taking the Scopus\footnote{\url{https://www.scopus.com}} indexation as a criterion. 
The research productivity of an actor is measured by counting the number of publications available in Scopus \citep{hancean2021coauthorship}. The data comprises all papers of each of the 3,600 seed authors and all papers published by any of their coauthors -- including those papers published by coauthors without any of the seed authors -- where the year of publication is from 2007 to 2019. The number of authors of any paper is limited to 100 (if a paper exceeds this limit, only the first 100 authors are considered), a constraint which is imposed by the data collection from Scopus. Additional covariates (e.\,g., sex, age, or education) for authors are not available in the Scopus dataset and are therefore not considered in our analysis. We will discuss below that the lack of controlling for covariates might distort some findings on network effects. We emphasize that the empirical analysis given in this paper is for illustrating the use of RHEM and RHOM in coauthor networks -- rather than for drawing empirical conclusions. The distributions (histograms) of the number of authors per paper is shown in Fig.~\ref{fig:coauthor_sizes} and descriptive statistics of the three coauthor networks are given in Table~\ref{tab:descriptive_statistics}. The distributions suggest that there are relatively few papers affected by the limit on the number of authors per paper (more precisely, there are 297 papers reaching or exceeding 100 authors in physics, 128 in medicine, and 60 in the social sciences). 

\begin{table}
  \caption{Descriptive statistics of the empirical coauthor networks in the three disciplines: number of papers (i.\,e., publication events), number of unique authors, number of links in the two-mode author-paper networks, maximum and mean of the number of authors per paper (note that the number of authors per paper has been limited to 100; see the text for additional explanation and see Fig.~\ref{fig:coauthor_sizes} for the distribution of the number of authors per paper) and the number of papers per author.}
  \label{tab:descriptive_statistics}
  \begin{center}
\begin{tabular}{l@{\quad}r@{\quad}r@{\quad}r}
\hline
 & Physics & Medicine & Social Science \\
\hline
\#papers & 193,931 & 39,862 & 122,184 \\
\#authors & 388,182 & 145,942 & 284,766 \\
\#author-paper links & 1,851,677 & 494,724 & 1,008,680 \\
max \#authors per paper & 100 & 100 & 100 \\
mean \#authors per paper & 9.55 & 12.41 & 8.26 \\
max \#papers per author & 691 & 376 & 471 \\
mean \#papers per author & 4.77 & 3.39 & 3.54\\
\hline
\end{tabular} 
  \end{center}
\end{table}

\subsection{Results and discussion}
\label{sec:results}

We fit two types of models, RHEM explaining publication rates of hyperedges, defined in Sect.~\ref{sec:rhem}, and RHOM explaining the impact of published papers, defined in Sect.~\ref{sec:rom}. Both model families are specified with the statistics defined in Sect.~\ref{sec:statistics}. For both model families we estimate four models, one for each of the three disciplines and one joint model that estimates a single parameter vector to maximize the joint likelihood function, which is the product of the three discipline-specific likelihoods.

\paragraph{Explaining publication events (RHEM).} The models explaining publication rates associated with hyperedges are Cox proportional hazard models whose likelihood is given in Eq.~(\ref{eq:likelihood}). As we discussed above, we condition the risk sets to those hyperedges that contain the same number of authors as the observed publication events and we apply case-control sampling to ensure computational tractability, where we sample 10 controls (i.\,e., non-event hyperedges) for each observed event. As discussed in Sect.~\ref{sec:rhem}, results from previous work on the reliability of REM parameters estimated from sampled likelihoods \citep{lerner2020reliability} let us expect that the additional variation induced by sampling does not distort our findings qualitatively. We additionally checked the standard deviation of parameters from ten independent samples in one of the three disciplines and found that it is comparable with the standard errors.

The statistics for all observations (i.\,e., events and sampled non-events) are computed with an extension of the \textbf{eventnet} software\footnote{\url{https://github.com/juergenlerner/eventnet}} \citep{lerner2020reliability} to hyperevents. The correlation of explanatory variables (hyperedge statistics) is given in Table~\ref{tab:cor_statistics}. The highest correlations are among subset repetition of order one and two ($0.69$), subset repetition of order two and three ($0.78$), and subset repetition of order three and closure ($0.64$). The pairwise correlation between the subset repetition statistics and closure on the one hand and the prior success indicators on the other hand is almost non-existent (not exceeding $0.05$ in absolute value). The prior success statistics have rather small correlation among themselves, with a moderate exception among prior dyadwise shared success ($prior.succ^{(2)}$) and prior shared success among triads ($prior.succ^{(3)}$) which reach a correlation of $0.56$. The prior success indicators have rather small correlation with success disparity. In summary, correlation among explanatory variables seems to be sufficiently low for reliable parameter estimation, considering that we fit models to hundreds of thousands of observed events. On the other hand, it is rather obvious -- and can also be expected from a theoretical perspective -- that inclusion or exclusion of some statistics can change estimated parameters associated with some other statistics. For instance, in an additional analysis (not reported in this paper) we found that failure to control for subset repetition of order one flips the sign of subset repetition of order two. We can argue that if we want to assess the effect of prior \emph{shared} activity, then we should also control for prior \emph{individual} activity -- so that the model without subset repetition of order one should be discarded by theoretical arguments.
  
\begin{table}
  \caption{Correlation of hyperedge statistics computed from the combined observations (events and sampled controls) from the three disciplines.}
  \label{tab:cor_statistics}
  \begin{center}
\begin{tabular}{l@{\quad}r@{\quad}r@{\quad}r@{\quad}r@{\quad}r@{\quad}r@{\quad}r}
  \hline
 & $s.r^{(2)}$ & $s.r^{(3)}$ & $closure$ & $p.s^{(1)}$ & $p.s^{(2)}$ & $p.s^{(3)}$ & $succ.disp$ \\ 
  \hline
  $sub.rep^{(1)}$    & 0.69 & 0.47 & 0.23 & -0.05 & 0.02 & 0.03 & 0.51 \\ 
  $sub.rep^{(2)}$    & $\cdot$ & 0.78 & 0.47 & -0.03 & -0.00 & 0.01 & 0.23 \\ 
  $sub.rep^{(3)}$    & $\cdot$ & $\cdot$ & 0.64 & -0.02 & -0.00 & 0.00 & 0.14 \\ 
  $closure$         & $\cdot$ & $\cdot$ & $\cdot$ & -0.00 & 0.00 & 0.00 & 0.02 \\ 
  $prior.succ^{(1)}$ & $\cdot$ & $\cdot$ & $\cdot$ & $\cdot$ & 0.11 & 0.09 & 0.20 \\ 
  $prior.succ^{(2)}$ & $\cdot$ & $\cdot$ & $\cdot$ & $\cdot$ & $\cdot$ & 0.56 & 0.06 \\ 
  $prior.succ^{(3)}$ & $\cdot$ & $\cdot$ & $\cdot$ & $\cdot$ & $\cdot$ & $\cdot$ & 0.05 \\ 
   \hline
\end{tabular}
  \end{center}
\end{table}

  Before fitting the models, each statistic is standardized to mean equal to zero and standard deviation equal to one. Given the statistics of all observations, parameters are estimated with the \texttt{coxph} function in the R-package \textbf{survival}\footnote{\url{https://CRAN.R-project.org/package=survival}} \citep{therneau2013modeling}, which also applies to tied event times. Estimated parameters, standard errors, and significance levels are given in Table~\ref{tab:assembly_by_disciplines}.

\begin{table}
  \caption{CoxPH model for publication rates associated with hyperedges. Estimated parameters and standard errors (in brackets).}
  \label{tab:assembly_by_disciplines}
  \begin{center}
\begin{tabular}{l@{\quad}r@{\quad}r@{\quad}r@{\quad}r}
\hline
 & joint model & Physics & Medicine & Social Science \\
\hline
$sub.rep^{(1)}$     & $0.322 \; (0.000)^{***}$  & $0.278 \; (0.001)^{***}$  & $0.314 \; (0.003)^{***}$  & $0.368 \; (0.001)^{***}$  \\
$sub.rep^{(2)}$     & $-0.036 \; (0.000)^{***}$ & $-0.024 \; (0.000)^{***}$ & $-0.024 \; (0.001)^{***}$ & $0.003 \; (0.001)^{**\phantom{*}}$   \\
$sub.rep^{(3)}$     & $0.048 \; (0.000)^{***}$  & $0.049 \; (0.001)^{***}$  & $0.083 \; (0.001)^{***}$  & $0.076 \; (0.001)^{***}$  \\
$closure$                    & $-0.042 \; (0.001)^{***}$ & $-0.038 \; (0.001)^{***}$ & $-0.122 \; (0.005)^{***}$ & $-2.596 \; (0.044)^{***}$ \\
$prior.succ^{(1)}$  & $-0.222 \; (0.002)^{***}$ & $-0.714 \; (0.006)^{***}$ & $-0.795 \; (0.014)^{***}$ & $-0.126 \; (0.003)^{***}$ \\
$prior.succ^{(2)}$  & $0.027 \; (0.001)^{***}$  & $0.035 \; (0.001)^{***}$  & $0.017 \; (0.002)^{***}$  & $0.112 \; (0.002)^{***}$  \\
$prior.succ^{(3)}$  & $0.032 \; (0.001)^{***}$  & $0.043 \; (0.001)^{***}$  & $0.048 \; (0.002)^{***}$  & $0.106 \; (0.002)^{***}$  \\
$succ.disp.$ & $0.088 \; (0.000)^{***}$  & $0.201 \; (0.001)^{***}$  & $0.290 \; (0.004)^{***}$  & $0.075 \; (0.001)^{***}$  \\
\hline
AIC                        & 7,878,538.597               & 4,379,396.644               & 794,665.437                & 2,663,494.601               \\
Num. events                & 355,977                    & 193,931                    & 39,862                     & 122,184                    \\
Num. obs.                  & 3,675,733                   & 1,998,279                   & 419,075                    & 1,258,379                   \\
\hline
\multicolumn{5}{l}{\scriptsize{$^{***}p<0.001$, $^{**}p<0.01$, $^*p<0.05$}}
\end{tabular} 
  \end{center}
\end{table}

We find that the parameters associated with subset repetition of order one are consistently positive in all three disciplines and also in the joint model. This means that scientists who have published more in the past are more likely to be included in the set of authors of future papers. This is consistent with a rich-get-richer effect, found in previous research \citep{kronegger2012collaboration}, predicting that scientists with higher numbers of publications accumulate future publications at a higher rate.

The effect of familiarity among scientists (that is, a history of coauthoring) on future co-publication depends on the size of subsets of scientists having a history of collaboration. The parameters associated with subset repetition of order three are all positive, implying that hyperedges composed of triads of scientists who have co-published before are more likely to experience publication events in the future. In contrast, the effect of subset repetition of order two is negative in the physics and medicine coauthor network, but positive in social science. We note that subset repetition effects are nested within each other: hyperedges taking high values in $sub.rep^{(3)}$ necessarily also take higher values in $sub.rep^{(2)}$ and $sub.rep^{(1)}$. This could explain alternating signs on the associated parameters: it could be the case that subset repetition of order three overestimates a familiarity effect which is then corrected by the negative sign of subset repetition of order two -- which in turn is corrected by a positive sign associated with subset repetition of order one. The finding on alternating signs associated with nested statistics resembles a historic finding on alternating signs associated with $k$-star statistics in ERGM \citep{robins2007recent} -- an insight that eventually led to the development of new statistics (alternating $k$-star statistics or geometrically weighted degrees). Future work might consider the possibility of developing similarly scaled statistics, replacing subset repetition in RHEM.

We consistently find a negative closure effect, similar to previous results with RHEM \citep{lerner2021dynamic}. A negative closure effect in coauthor networks might seem surprising, given that previous research reported positive triadic closure \citep{ferligoj2015scientific}. We argue that their findings do not contradict ours since the results from \citet{ferligoj2015scientific} have been obtained by analyzing one-mode projections in which two scientists are linked if they coauthored a paper -- while our models estimate publication rates associated with hyperedges, that is, groups of scientists of any size. We discussed already in Sect.~\ref{sec:statistics} (also compare Fig.~\ref{fig:closure}) that closure in hyperevent networks has to be interpreted jointly with the size of hyperevents and with subset repetition of order two or higher. Large events (that is, publications with many authors) already produce a high number of closed triangles in the one-mode projection and the tendency to repeat collaborations reinforces these dense local clusters. Once we control for subset repetition, \emph{positive} closure would reveal a tendency to establish \emph{new} collaborations among scientists who are indirectly linked -- an example of such a closure effect would be provided by an event on the hyperedge $h'=\{D,E,H\}$ in Fig.~\ref{fig:closure}. We discussed above that such a positive closure effect would imply that overlapping dense clusters have a tendency to merge over time. In contrast, the \emph{negative} closure effect which we found in our empirical analysis -- together with positive subset repetition -- implies the emergence and maintenance of stable, overlapping, but non-merging, dense clusters. Some scientists have bridging positions between dense groups but their coauthors do not necessarily start collaborating among themselves -- preventing that groups merge. To further corroborate our interpretation of the negative closure effect, we created a simulated, artificial dataset of 1,000 randomly generated hyperevents on a set of 19 actors that give rise to two dense groups, each of size ten, that are overlapping in one actor (the ``broker''). Events happen only within groups, so that the artificial network has a planted structure of two dense, overlapping but non-merging groups. Fitting RHEM to this artificial network indeed yields a negative closure effect, if we control for prior shared activity via subset repetition of order two and three. Thus, findings on this artificial data are consistent with our interpretation of the negative closure effect.

While effects discussed so far modeled publication rates dependent on the \emph{number} of previous (individual or joint) publications, the following four effects explain co-publication by the \emph{impact} of previous publications. Results are qualitatively very consistent over the three disciplines. We find that prior success of individual scientists has a negative effect on their participation in future publication events. This could point to a saturation effect where successful individuals feel less urge to publish -- everything else being equal. On the other hand, this effect also has to be interpreted jointly with prior success of order two and three: similar as for subset repetition, the different statistics based on prior success are nested within each other.

In contrast to prior individual success, prior \emph{shared} success has a consistently positive effect on the probability to co-publish again in the future. All parameters associated with prior success of order two and three are positive in all three disciplines and in the joint model. This means that pairs (or triads) of authors whose previous joint publications had a higher impact are more inclined to continue collaboration.

Finally, prior success disparity, that is, the variation of prior success among the members of hyperedges, has a consistently positive effect on future publication rates. This means that hyperedges composed of a mixture of formerly successful and unsuccessful scientists are more likely to publish than hyperedges consisting only of successful, or only of unsuccessful scientists.

To assess the explanatory power of the fitted RHEM we follow the approach from \citet{lerner2021dynamic} and compare the predicted rate on the hyperedges of observed events with the predicted rate on controls (i.\,e., non-event hyperedges) sampled from the risk set at the event time. We find that the median percentile of the rate of observed events in the distribution of the predicted rate of the associated controls is 0.976. This means that the predicted rate of half of the observed events exceeds the predicted rate of more than 97\% of all associated controls sampled at the event time. This finding implies that the fitted RHEM typically assigns relatively high rates to those hyperedges on which events are actually observed. We emphasize, however, that the task of identifying (or ``predicting'') the hyperedges of observed events in the entire risk set is virtually impossible, keeping in mind that the largest size-constrained risk sets in our analysis have some ${388,000 \choose 100}$ elements, which is by a very rough estimate approximately $10^{400}$. We consider it as an important topic for future work to develop further methods to assess the goodness of fit of RHEM. We discuss this aspect in more detail in the conclusion and future work section.

While conditioning on hyperedge size ensures that observed events and associated sampled non-event hyperedges are better comparable, our models still assume that effects are homogeneous over hyperedges of different sizes. For instance, the baseline publication rate on hyperedges of size ten is much higher than the baseline publication rate on hyperedges of size 20 (compare Fig.~\ref{fig:coauthor_sizes}). Yet the models estimated so far assume that given hyperedge statistics have a consistent effect for hyperedges of difference size. For instance, our models reveal that those hyperedges of size ten that have high values in the closure statistic are less likely to co-publish than hyperedges of size ten that have lower values in this statistic -- and that those hyperedges of size 20 that have high values in the closure statistic are also less likely to co-publish than hyperedges of size 20 that have lower values in this statistic. Thus, while the baseline rate varies by size, the \emph{change} in the relative event rate implied by a given network effect is assumed to be the same for small and for large hyperedges. Clearly, effects are not necessarily homogeneous over hyperedge size -- and it is straightforward to test or control for possible heterogeneity in RHEM or RHOM by interacting effects with the size of hyperedges. We illustrate this approach in the appendix.

\paragraph{Explaining the impact of published papers (RHOM).} The other family of models seeks to explain the impact (that is, the normalized number of citations) of published papers. Observations used in estimating these models are all observed events. (Non-events are not considered here since a hyperedge that did not publish a paper has no associated impact.) Estimation is done by ordinary least squares (linear regression), where the outcome variable is the normalized number of citations, defined in Eq.~(\ref{eq:impact}), and explanatory variables are the same statistics that we used in the models explaining publication rates (see Table~\ref{tab:assembly_by_disciplines}) plus one additional statistic, equal to the number of authors of the paper.\footnote{We note that the number of authors cannot be taken as an explanatory variable in the model explaining publication rates. Since we condition the risk sets on the hyperedge size of the observed publication events, any effect of the number of authors would cancel out in the factors of the likelihood function given in Eq.~(\ref{eq:likelihood}). This is a general property of the Cox proportional hazard model that has no intercept, since any offset in the event rate is absorbed into the baseline rate.} All explanatory variables are standardized (subtracting the mean and dividing by the standard deviation) but the outcome variable is not. This means that the estimated parameters can be interpreted as a change in the number of citations implied by an increase of one standard deviation in the explanatory variable. Results are reported in Table~\ref{tab:performance_by_disciplines}. 

\begin{table}
  \caption{OLS for impact (normalized number of citations) of published papers. Estimated parameters and standard errors (in brackets).}
  \label{tab:performance_by_disciplines}
  \begin{center}
\begin{tabular}{l@{\quad}r@{\quad}r@{\quad}r@{\quad}r}
\hline
 & joint model & Physics & Medicine & Social Science \\
\hline
(Intercept)                & $-0.000 \; (0.249)\phantom{^{***}}$       & $-0.231 \; (0.175)\phantom{^{***}}$       & $-0.945 \; (0.731)\phantom{^{***}}$      & $1.814 \; (0.953)^{+\phantom{**}}$   \\
$sub.rep^{(1)}$     & $-0.504 \; (0.388)\phantom{^{***}}$       & $1.645 \; (0.261)^{***}$  & $-3.449 \; (1.187)^{**\phantom{*}}$ & $-6.847 \; (1.203)^{***}$    \\
$sub.rep^{(2)}$     & $-1.796 \; (0.483)^{***}$ & $-0.703 \; (0.324)^{*\phantom{**}}$   & $0.901 \; (0.898)\phantom{^{***}}$       & $-4.228 \; (1.440)^{**\phantom{*}}$     \\
$sub.rep^{(3)}$     & $-1.959 \; (0.455)^{***}$ & $-1.777 \; (0.291)^{***}$ & $-2.912 \; (0.951)^{**\phantom{*}}$ & $-1.049 \; (1.473)\phantom{^{***}}$          \\
$closure$                    & $-0.228 \; (0.335)\phantom{^{***}}$       & $-0.694 \; (0.181)^{***}$ & $-0.902 \; (2.751)\phantom{^{***}}$      & $24.683 \; (13.856)^{+\phantom{*}}$ \\
$prior.succ^{(1)}$  & $14.610 \; (0.323)^{***}$ & $15.227 \; (0.357)^{***}$ & $5.483 \; (0.574)^{***}$ & $16.970 \; (0.669)^{***}$    \\
$prior.succ^{(2)}$  & $4.739 \; (0.417)^{***}$  & $1.613 \; (0.382)^{***}$  & $7.757 \; (0.891)^{***}$ & $4.307 \; (0.842)^{***}$     \\
$prior.succ^{(3)}$  & $2.602 \; (0.371)^{***}$  & $9.505 \; (0.341)^{***}$  & $1.738 \; (0.874)^{*\phantom{**}}$   & $-0.437 \; (0.722)\phantom{^{***}}$          \\
$succ.disp.$ & $-3.817 \; (0.284)^{***}$ & $-4.267 \; (0.291)^{***}$ & $4.678 \; (1.435)^{**\phantom{*}}$  & $-2.536 \; (0.634)^{***}$    \\
$num.auth.$        & $8.424 \; (0.271)^{***}$  & $6.103 \; (0.179)^{***}$  & $5.831 \; (0.429)^{***}$ & $16.276 \; (0.922)^{***}$    \\
\hline
R$^2$                      & 0.023                     & 0.057                     & 0.029                    & 0.020                        \\
Num. obs.                  & 355,977                    & 193,931                    & 39,862                    & 122,184                       \\
RMSE                       & 148.666                   & 73.153                    & 105.164                  & 228.296                      \\
\hline
\multicolumn{5}{l}{\scriptsize{$^{***}p<0.001$, $^{**}p<0.01$, $^*p<0.05$, $^{+}p<0.1$}}
\end{tabular} 
  \end{center}
\end{table}

With few exceptions we find that subset repetition of order one, two, and three is rather negative for the impact of papers. This means that papers written by more active individuals and papers written by pairs and triads of scientists with more previous joint publications tend to gather lower numbers of citations -- all other things being equal (note that previous \emph{successful} publications have an increasing effect on future impact, see below). The only significant effect going in the other direction is $sub.rep^{(1)}$ which is positive in the physics network. Three parameters (out of 12) in this family are not significant.

We find no consistent effect of closure on the impact of papers. The associated parameter is significantly negative in the physics network, significantly positive (albeit only at the 10\% level) in the social science network, and non-significant in medicine.

In contrast to these rather inconsistent effects, modeling the impact of papers by the \emph{number} of previous (joint) publications of its authors, we find consistently positive effects of prior individual and shared \emph{success} on future impact. All but one parameter associated with prior success of order one, two, and three are significantly positive in all three disciplines. Thus, teams composed of scientists that have a history of prior individual or shared success tend to write more successful papers in the future.

Prior success disparity (i.\,e., the standard deviation over the prior success of team members) has a significantly negative effect on future impact in the physics and social science networks. This means that in these two disciplines teams mixing previously successful scientists with previously unsuccessful scientists are less successful than teams that are more homogeneous with respect to prior success, consistent with findings from \citep{ahmadpoor2019decoding}. Results have to be interpreted together with the positive effect of the average prior individual or shared success: if two teams have identical values in prior success of order one, two, and three, then a team in which this past success is distributed more evenly is more successful in the future than a team that concentrates prior success in a subset.
However, prior success disparity has a significantly positive effect in medicine, implying that -- all other things being equal -- high impact papers in medicine tend to be written by teams composed of highly successful and less successful authors. This finding could point to differences in publication dynamics across disciplines.

We find that the number of authors has a positive effect on the impact of papers in all three disciplines. This is consistent with previous findings that larger teams have a higher probability to write high-impact papers \citep{wuchty2007increasing}.

Overall, the variance explained by the model is rather small ($R^2$ equal to 0.023). We explain this observation by arguing that predicting the correct number of citations of a paper is an intrinsically difficult task -- and also by the skewed distribution of the number of citations. In fact, RHOM as we specify in this paper, are intended to assess whether given characteristics of sets of authors (operationalized by hyperedge statistics) tend to increase or decrease the expected number of citations gathered by their papers. The largest effect sizes that we found are for prior individual success and reach values up to 16, implying that an increase in prior individual success by one standard deviation tends to increase the expected number of citations by 16. Thus, effects could be considered meaningful from the point of view of their implications for the number of citations. However, predicting the true number of citations gathered by the individual papers with a small residual error seems to be a difficult task -- at least for the models considered in this paper.

It is informative to compare effects in the model explaining publication rates with effects explaining impact of published papers. Do mechanisms explaining team assembly have a tendency to produce high impact papers? Answers to this question depend on the type of effect. The effect of prior \emph{shared} success (that is, prior success of order two or three) is consistently positive in the model for publication rates and in the model explaining impact of published papers. That is, scientists are inclined to publish with those with which they have a history of successful collaboration -- and this effect is likely to produce future papers of high impact. The situation is different for prior \emph{individual} success. The team assembly model (Table~\ref{tab:assembly_by_disciplines}) suggests that scientists who were individually successful typically publish less in the future (all other things being equal) -- but if they do, they are likely to produce high impact papers, again. The effect of prior success disparity goes in the opposite direction in the team assembly and team performance models in the physics and social science networks. Diverse teams (being a mixture of successful and unsuccessful scientist) publish more but typically produce lower impact per paper. We also find some inconsistent effects of subset repetition in the two models. Past individual publication activity increases the likelihood to be included among the authors of future publications (preferential attachment effect) -- but typically produces papers of lower impact.

\section{Conclusion, limitations, and future work}
\label{sec:conclusion}

We discussed, elaborated, and applied RHEM \citep{lerner2021dynamic} for analyzing scientific collaboration and proposed RHOM for modeling scientific impact in coauthor networks. RHEM can model publication rates associated with hyperedges (i.\,e., groups of scientists of any size), dependent on previous publications on the same or related hyperedges, as well as on covariates of individuals or groups. The related RHOM can model the impact of published papers (e.\,g., the normalized number of citations) dependent on the same explanatory variables as in models for the publication rate. This allows to compare patterns of scientific team formation with those explaining the impact of scientific collaboration. For instance, we could analyze whether factors that increase the likelihood of joint publications also tend to increase the expected impact of papers.

We illustratively applied RHEM and RHOM to empirical coauthor networks of scientists working in three disciplines. Our models explained publication rates -- or impact of published papers, respectively -- dependent on past individual activity, familiarity, triadic closure, prior individual and shared success, and prior success disparity. Many of our illustrative findings are consistent with previous findings reported in the literature -- yet we also provided refined insights into some of the network effects underlying the dynamics of coauthor networks. The most relevant empirical finding -- keeping in mind that our analysis is meant for illustration and not for drawing conclusive insights -- seems to be related with the interplay of subset repetition and closure effects in hyperevent networks. As discussed in Sects.~\ref{sec:statistics} and~\ref{sec:results} (also compare Fig.~\ref{fig:closure}), an over-representation of closed triangles in one-mode coauthor networks can be explained by papers with many authors and by a tendency to repeat previous collaborations. Controlling for these effects, we found a \emph{negative} effect of closure on the publication rate. We discussed above that positive subset repetition and negative closure effects can explain the emergence and maintenance of overlapping yet stable dense subgroups. Some scientists occupy bridging positions between dense groups -- but their coauthors do not necessarily start collaborating among themselves. In contrast, a positive closure effect would imply that overlapping dense groups have a tendency to merge over time. Analyzing and testing this interpretation more thoroughly is a promising direction for future work.

We further recall that the empirical analysis given in this paper is meant for illustrating the use of RHEM and RHOM, rather than for drawing conclusive insights. It is possible that the lack of exogenous covariates (e.\,g., institutional affiliation or joint membership in research projects) in our analysis might distort empirical findings. Notably the interpretation of some effects in RHEM or RHOM (e.\,g., subset repetition of order two or higher or closure) is contingent on an assumed underlying clustering into dense groups. Since such groups can also be explained, for instance, by the aforementioned covariates (affiliation or membership in research projects), it is plausible that inclusion of such covariate effects might change empirical network effects. We note that covariate effects can be included in RHEM and RHOM as indicated in Sect.~\ref{sec:statistics}, if information on covariates is available.
 
There are also several methodological aspects of our model that can be refined or improved. First, we assume conditional independence of simultaneous publication events. Both our model families, RHEM and RHOM, can express that publication events -- or the resulting impact of papers -- with a given time stamp depend on papers published earlier. On the other hand, we assume that simultaneous events, i.\,e., papers with the same publication time, are conditionally independent of each other, given the networks of previous events. Previous work has shown that the validity of the assumption of conditional independence depends on the spacing of observations \citep{lerner2013conditional}. If time granularity is fine enough, models assuming conditional independence yield approximately the same results as more sophisticated models that account for mutually dependent dyadic observations. It has to be addressed in future work whether assuming conditional independence in RHEM or RHOM for coauthor networks distorts estimated network effects, whether coding publication events with more fine-grained time stamps improves model validity, and/or whether models have to be augmented to cope with mutual dependence among simultaneous events.

A second modeling aspect that has to be addressed in future work is the handling of the impact of papers. In our empirical analysis we quantified the impact of a published paper by the normalized number of citations that the paper gathers by the end of the observation period. Our models assumed that the impact of a paper is transparent to others in the year following the paper's year of publication. For instance, the impact of a paper published in year $t$ is assumed to be known in year $t+1$. Clearly, this cannot be true in the direct sense since some of the citing papers are published later than in the year $t+1$. Indeed, our models take the normalized number of citations as a proxy for the paper's impact -- and assumed that other scientists recognize this impact by different signals (which might include their own assessment of the quality of the paper, talks given by the authors, qualitative reactions from the research community, or the number of citations known in the year $t+1$). These assumptions need to be tested for validity -- or the impact of papers has to be modeled in different ways. One possibility would be not to use the cumulative number of citations at the end of the observation period, but rather to model individual citation events as fine-grained signals of scientific impact. This could be achieved by a joint modeling of coauthor networks and citation networks. Citation networks can be regarded as networks of \emph{directed} relational hyperevents: each published paper gives rise to an event in which the citing paper (source of the event) cites any number of previously published papers (set of targets of the event). This opens the possibility to analyze the co-evolution of coauthoring and citation networks. Models explaining incoming citation events would then provide a more fine-grained analysis of the impact of scientific papers. Directed relational hyperevents can in principle also be modeled with RHEM \citep{lerner2019rem} -- although to the best of our knowledge directed RHEM have not yet been applied to empirical data. A different, related model has been proposed by \citep{mulder2021latent}.

Another topic for future methodological contributions is to address whether there are alternatives to conditioning the size of non-event hyperedges in RHEM. \citet{lerner2021dynamic} advocate conditional-size RHEM, but have also compared their results to RHEM fitted with unconstrained risk sets. The situation is somewhat clearer in our case: while \citet{lerner2021dynamic} analyzed RHEM on a sample of up to 22 actors, the number of nodes in our empirical networks is as high as 388,000. Drawing non-event hyperedges from the unconstrained risk set would lead to hyperedges containing on average almost 200,000 scientists -- an absurd number of coauthors. Yet there might be alternatives to imposing the hard constraint on hyperedge size. It could be possible to develop sampling strategies that \emph{control for} the observed number of participants in hyperevents, in the sense that the expected size matches the observed size, rather than condition upon. Such an approach would require that RHEM include the hyperedge size (number of authors in our case) as an explanatory variable, most likely associated with a negative parameter, and draw non-event hyperedges proportional to their probability implied by the model. This would loosely resemble established methods to sample from ERGM \citep{lkr-ergmsn-13}, where typically networks are sampled controlling for the number of edges -- as an alternative to condition on the exact number of observed edges. However, such sampling algorithms are likely to be more complicated than uniform sampling from the size-constrained risk sets, which might entail an increased computational runtime. Moreover, it is unclear how to incorporate the benefits of nested case-control sampling into such more sophisticated sampling algorithms. Note that the motivation for case-control sampling is the insight that events typically carry more valuable information than non-events \citep{bgl-mascdcphm-95}, motivating the over-representation of observed events over controls.

Last but not least, developing better methods to assess the goodness of fit of estimated RHEM is an important direction for future work. In our paper we only scratched this aspect by comparing the predicted rate of observed hyperevents with the predicted rate of alternative hyperedges (``controls'') sampled from the risk set at the event time. In general, we found that the fitted RHEM typically assign relatively high rates to those hyperedges on which events are actually observed. We emphasize, however, that the task of identifying (or ``predicting'') the hyperedges of observed events in the entire risk set is virtually impossible, due to the sheer size of the risk set. A possible way to circumvent these problems is to assess the goodness of fit not by whether a RHEM can predict the exact hyperedges of observed events -- but rather by comparing the distributions of hyperedge statistics over predicted events with those over the observed events. This approach, which mimics established goodness of fit procedures for ERGM \citep{lkr-ergmsn-13}, would not check whether a RHEM is able to identify or predict the exact events but -- more realistically -- whether it succeeds in generating the structural characteristics of observed events. 

We mentioned already in the introduction that RHEM and RHOM are not restricted to scientific networks but can be applied to other situations in which relational hyperevents represent a team that tackles a given task, provides a service, or produces a product and where these events are associated with a measurable outcome. Besides scientific collaboration, we mentioned as other potential application areas project teams in companies developing a new product or registering a patent \citep{trajtenberg1990penny}, teams of artists and other staff producing a movie \citep{ravasz2003hierarchical}, sports teams \citep{mukherjee2019prior}, or medical teams performing a given surgery \citep{pallotti2020lost}. While technically, RHEM and RHOM could be applied to these empirical applications almost unchanged (given the required data), each scenario might come with additional constraints that have to be considered in the models. For instance, employees in a corporate environment are typically not free in assembling into project teams with the partners of their choice. As a concrete example, surgery teams are most likely constrained by working hours, since only staff working in the same shift can form a team, and by requirements on the roles or positions of team members (e.\,g., a prescribed number of surgeons, assistants, anesthetists, or nurses). Such constraints can be incorporated into RHEM by defining the risk set accordingly. In contrast, scientific team formation, or the selection of coauthors, is likely to be less affected by such hard constraints.

\paragraph{Acknowledgment.} Financially supported by Deutsche Forschungsgemeinschaft (DFG) -- 321869138 and Executive Agency for Higher Education, Research, Development and Innovation Funding (UEFISCDI grant, code PN-III-P4-ID-PCE-2020-2828).
  

\begin{thebibliography}{}

\bibitem[\protect\citename{Ahmadpoor \& Jones, }2019]{ahmadpoor2019decoding}
Ahmadpoor, Mohammad, \& Jones, Benjamin~F. (2019).
\newblock Decoding team and individual impact in science and invention.
\newblock {\em Proceedings of the national academy of sciences},  201812341.

\bibitem[\protect\citename{Berge, }1989]{berge1989hypergraphs}
Berge, Claude. (1989).
\newblock {\em Hypergraphs: combinatorics of finite sets}.
\newblock North-Holland.

\bibitem[\protect\citename{Blaschke {\em et~al.}\relax,
  }2012]{blaschke2012organizations}
Blaschke, Steffen, Schoeneborn, Dennis, \& Seidl, David. (2012).
\newblock Organizations as networks of communication episodes: Turning the
  network perspective inside out.
\newblock {\em Organization studies}, {\bf 33}(7), 879--906.

\bibitem[\protect\citename{Borgan {\em et~al.}\relax, }1995]{bgl-mascdcphm-95}
Borgan, {\O}, Goldstein, L, \& Langholz, B. (1995).
\newblock Methods for the analysis of sampled cohort data in the {C}ox
  proportional hazards model.
\newblock {\em The annals of statistics},  1749--1778.

\bibitem[\protect\citename{B{\"o}rner {\em et~al.}\relax,
  }2010]{borner2010multi}
B{\"o}rner, Katy, Contractor, Noshir, Falk-Krzesinski, Holly~J, Fiore,
  Stephen~M, Hall, Kara~L, Keyton, Joann, Spring, Bonnie, Stokols, Daniel,
  Trochim, William, \& Uzzi, Brian. (2010).
\newblock A multi-level systems perspective for the science of team science.
\newblock {\em Science translational medicine}, {\bf 2}(49), 49cm24--49cm24.

\bibitem[\protect\citename{Brandes {\em et~al.}\relax, }2009]{bls-ness-09}
Brandes, Ulrik, Lerner, J{\"u}rgen, \& Snijders, Tom~{A.\ B.} (2009).
\newblock Networks evolving step by step: Statistical analysis of dyadic event
  data.
\newblock {\em Pages  200--205 of:} {\em Proc.\ 2009 intl.\ conf.\ advances in
  social network analysis and mining (asonam)}.
\newblock IEEE.

\bibitem[\protect\citename{Breiger, }1974]{breiger1974duality}
Breiger, Ronald~L. (1974).
\newblock The duality of persons and groups.
\newblock {\em Social forces}, {\bf 53}(2), 181--190.

\bibitem[\protect\citename{Bretto, }2013]{bretto2013hypergraph}
Bretto, Alain. (2013).
\newblock {\em Hypergraph theory}.
\newblock Cham: Springer.

\bibitem[\protect\citename{Butts, }2008]{b-refsa-08}
Butts, Carter~T.\. (2008).
\newblock A relational event framework for social action.
\newblock {\em Sociological methodology}, {\bf 38}(1), 155--200.

\bibitem[\protect\citename{Chodrow \& Mellor, }2020]{chodrow2020annotated}
Chodrow, Philip, \& Mellor, Andrew. (2020).
\newblock Annotated hypergraphs: models and applications.
\newblock {\em Applied network science}, {\bf 5}(1), 9.

\bibitem[\protect\citename{Chodrow, }2019]{chodrow2019configuration}
Chodrow, Philip~S. (2019).
\newblock Configuration models of random hypergraphs and their applications.
\newblock {\em arxiv preprint arxiv:1902.09302}.

\bibitem[\protect\citename{Cox, }1972]{cox1972regression}
Cox, David. (1972).
\newblock Regression models and life-tables.
\newblock {\em Journal of the royal statistical society. series b
  (methodological)}, {\bf 34}(2), 87--22.

\bibitem[\protect\citename{Cugmas {\em et~al.}\relax,
  }2017]{cugmas2019scientific}
Cugmas, Marjan, Ferligoj, Anu{\v{s}}ka, \& Kronegger, Luka. (2017).
\newblock Scientific co-authorship networks.
\newblock {\em Chap. 14, pages  251--275 of:} Doreian, Patrick, Batagelj,
  Vladimir, \& Ferligoj, Anuska (eds), {\em Advances in network clustering and
  blockmodeling}.
\newblock John Wiley \& Sons.

\bibitem[\protect\citename{Davis {\em et~al.}\relax, }1941]{davis1941deep}
Davis, A., Gardner, B.B., \& Gardner, M.R. (1941).
\newblock {\em Deep south}.
\newblock The University of Chicago Press.

\bibitem[\protect\citename{Ferligoj {\em et~al.}\relax,
  }2015]{ferligoj2015scientific}
Ferligoj, Anu{\v{s}}ka, Kronegger, Luka, Mali, Franc, Snijders, Tom~AB, \&
  Doreian, Patrick. (2015).
\newblock Scientific collaboration dynamics in a national scientific system.
\newblock {\em Scientometrics}, {\bf 104}(3), 985--1012.

\bibitem[\protect\citename{Freeman, }2003]{freeman2003finding}
Freeman, Linton~C. (2003).
\newblock Finding social groups: A meta-analysis of the {S}outhern {W}omen
  data.
\newblock {\em Pages  39--97 of:} Breiger, R.\, Carley, C.\, \& Pattison, P.
  (eds), {\em Dynamic social network modeling and analysis: Workshop summary
  and papers}.
\newblock Washington, DC: National Research Council, The National Academies
  Press.

\bibitem[\protect\citename{Guimera {\em et~al.}\relax, }2005]{guimera2005team}
Guimera, Roger, Uzzi, Brian, Spiro, Jarrett, \& Amaral, Luis A~Nunes. (2005).
\newblock Team assembly mechanisms determine collaboration network structure
  and team performance.
\newblock {\em Science}, {\bf 308}(5722), 697--702.

\bibitem[\protect\citename{H{\^a}ncean \& Perc, }2016]{hancean2016homophily}
H{\^a}ncean, Marian-Gabriel, \& Perc, Matja{\v{z}}. (2016).
\newblock Homophily in coauthorship networks of east european sociologists.
\newblock {\em Scientific reports}, {\bf 6}, 36152.

\bibitem[\protect\citename{H{\^a}ncean {\em et~al.}\relax,
  }2021]{hancean2021coauthorship}
H{\^a}ncean, Marian-Gabriel, Perc, Matja{\v{z}}, \& Lerner, J{\"u}rgen. (2021).
\newblock The coauthorship networks of the most productive european
  researchers.
\newblock {\em Scientometrics}, {\bf 126}(1), 201--224.

\bibitem[\protect\citename{Hayat {\em et~al.}\relax,
  }2020]{hayat2020differential}
Hayat, Tsahi, Dimitrova, Dimitrina, \& Wellman, Barry. (2020).
\newblock The differential impact of network connectedness and size on
  researchers' productivity and influence.
\newblock {\em Information, communication \& society},  1--18.

\bibitem[\protect\citename{Koskinen \& Edling, }2012]{koskinen2012modelling}
Koskinen, Johan, \& Edling, Christofer. (2012).
\newblock Modelling the evolution of a bipartite network---peer referral in
  interlocking directorates.
\newblock {\em Social networks}, {\bf 34}(3), 309--322.

\bibitem[\protect\citename{Krivitsky \& Handcock,
  }2014]{krivitsky2014separable}
Krivitsky, Pavel~N, \& Handcock, Mark~S. (2014).
\newblock A separable model for dynamic networks.
\newblock {\em Journal of the royal statistical society: Series b (statistical
  methodology)}, {\bf 76}(1), 29--46.

\bibitem[\protect\citename{Kronegger {\em et~al.}\relax,
  }2012]{kronegger2012collaboration}
Kronegger, Luka, Mali, Franc, Ferligoj, Anu{\v{s}}ka, \& Doreian, Patrick.
  (2012).
\newblock Collaboration structures in slovenian scientific communities.
\newblock {\em Scientometrics}, {\bf 90}(2), 631--647.

\bibitem[\protect\citename{Kumar, }2015]{kumar2015co}
Kumar, Sameer. (2015).
\newblock Co-authorship networks: a review of the literature.
\newblock {\em Aslib journal of information management}.

\bibitem[\protect\citename{Lerner \& Lomi, }2020]{lerner2020reliability}
Lerner, J{\"u}rgen, \& Lomi, Alessandro. (2020).
\newblock Reliability of relational event model estimates under sampling: how
  to fit a relational event model to 360 million dyadic events.
\newblock {\em Network science}, {\bf 8}(1), 97--135.

\bibitem[\protect\citename{Lerner {\em et~al.}\relax,
  }2013a]{lerner2013conditional}
Lerner, J{\"u}rgen, Indlekofer, Natalie, Nick, Bobo, \& Brandes, Ulrik.
  (2013a).
\newblock Conditional independence in dynamic networks.
\newblock {\em Journal of mathematical psychology}, {\bf 57}(6), 275--283.

\bibitem[\protect\citename{Lerner {\em et~al.}\relax, }2013b]{lbsb-mftien-13}
Lerner, J{\"u}rgen, Bussmann, Margit, Snijders, Tom~A.B., \& Brandes, Ulrik.
  (2013b).
\newblock Modeling frequency and type of interaction in event networks.
\newblock {\em Corvinus journal of sociology and social policy}, {\bf 4}(1),
  3--32.

\bibitem[\protect\citename{Lerner {\em et~al.}\relax, }2019]{lerner2019rem}
Lerner, J{\"u}rgen, Tranmer, Mark, Mowbray, John, \& Hancean, Marian-Gabriel.
  (2019).
\newblock {REM} beyond dyads: relational hyperevent models for multi-actor
  interaction networks.
\newblock {\em arxiv preprint arxiv:1912.07403}.
\newblock \texttt{https://arxiv.org/abs/1912.07403}.

\bibitem[\protect\citename{Lerner {\em et~al.}\relax, }2021]{lerner2021dynamic}
Lerner, J{\"u}rgen, Lomi, Alessandro, Mowbray, John, Rollings, Neil, \&
  Tranmer, Mark. (2021).
\newblock Dynamic network analysis of contact diaries.
\newblock {\em Social networks}, {\bf 66}, 224--236.

\bibitem[\protect\citename{Lusher {\em et~al.}\relax, }2013]{lkr-ergmsn-13}
Lusher, Dean, Koskinen, Johan, \& Robins, Garry (eds). (2013).
\newblock {\em Exponential random graph models for social networks}.
\newblock Cambridge University Press.

\bibitem[\protect\citename{Mukherjee {\em et~al.}\relax,
  }2019]{mukherjee2019prior}
Mukherjee, Satyam, Huang, Yun, Neidhardt, Julia, Uzzi, Brian, \& Contractor,
  Noshir. (2019).
\newblock Prior shared success predicts victory in team competitions.
\newblock {\em Nature human behaviour}, {\bf 3}(1), 74--81.

\bibitem[\protect\citename{Mulder \& Hoff, }2021]{mulder2021latent}
Mulder, Joris, \& Hoff, Peter~D. (2021).
\newblock A latent variable model for relational events with multiple
  receivers.
\newblock {\em arxiv preprint arxiv:2101.05135}.

\bibitem[\protect\citename{Pallotti {\em et~al.}\relax,
  }2020]{pallotti2020lost}
Pallotti, Francesca, Weldon, Sharon~Marie, \& Lomi, Alessandro. (2020).
\newblock Lost in translation: Collecting and coding data on social relations
  from audio-visual recordings.
\newblock {\em Social networks}.
\newblock forthcoming, \texttt{https://doi.org/10.1016/j.socnet.2020.02.006}.

\bibitem[\protect\citename{Perry \& Wolfe, }2013]{perry2013point}
Perry, Patrick~O, \& Wolfe, Patrick~J. (2013).
\newblock Point process modelling for directed interaction networks.
\newblock {\em Journal of the royal statistical society: Series b (statistical
  methodology)}, {\bf 75}(5), 821--849.

\bibitem[\protect\citename{Ravasz \& Barab{\'a}si,
  }2003]{ravasz2003hierarchical}
Ravasz, Erzs{\'e}bet, \& Barab{\'a}si, Albert-L{\'a}szl{\'o}. (2003).
\newblock Hierarchical organization in complex networks.
\newblock {\em Physical review e}, {\bf 67}(2), 026112.

\bibitem[\protect\citename{Robins {\em et~al.}\relax, }2007]{robins2007recent}
Robins, Garry, Snijders, Tom, Wang, Peng, Handcock, Mark, \& Pattison,
  Philippa. (2007).
\newblock Recent developments in exponential random graph (p*) models for
  social networks.
\newblock {\em Social networks}, {\bf 29}(2), 192--215.

\bibitem[\protect\citename{Seidman, }1981]{seidman1981structures}
Seidman, Stephen~B. (1981).
\newblock Structures induced by collections of subsets: A hypergraph approach.
\newblock {\em Mathematical social sciences}, {\bf 1}(4), 381--396.

\bibitem[\protect\citename{Snijders, }2005]{s-mlnd-05}
Snijders, Tom~A.B.\. (2005).
\newblock Models for longitudinal network data.
\newblock  Carrington, Peter~J.\, Scott, John, \& Wasserman, Stanley (eds),
  {\em Models and methods in social network analysis}.
\newblock Cambridge University Press.

\bibitem[\protect\citename{Snijders {\em et~al.}\relax,
  }2013]{snijders2013model}
Snijders, Tom~AB, Lomi, Alessandro, \& Torl{\'o}, Vanina~Jasmine. (2013).
\newblock A model for the multiplex dynamics of two-mode and one-mode networks,
  with an application to employment preference, friendship, and advice.
\newblock {\em Social networks}, {\bf 35}(2), 265--276.

\bibitem[\protect\citename{Stadtfeld \& Block, }2017]{sb-iat-17}
Stadtfeld, Christoph, \& Block, Per. (2017).
\newblock Interactions, actors, and time: Dynamic network actor models for
  relational events.
\newblock {\em Sociological science}, {\bf 4}, 318--352.

\bibitem[\protect\citename{Therneau \& Grambsch, }2013]{therneau2013modeling}
Therneau, Terry~M, \& Grambsch, Patricia~M. (2013).
\newblock {\em Modeling survival data: extending the {C}ox model}.
\newblock Springer Science \& Business Media.

\bibitem[\protect\citename{Trajtenberg, }1990]{trajtenberg1990penny}
Trajtenberg, Manuel. (1990).
\newblock A penny for your quotes: patent citations and the value of
  innovations.
\newblock {\em The rand journal of economics},  172--187.

\bibitem[\protect\citename{Vu {\em et~al.}\relax, }2015]{vpr-remslm-15}
Vu, Duy, Pattison, Philippa, \& Robins, Garry. (2015).
\newblock Relational event models for social learning in moocs.
\newblock {\em Social networks}, {\bf 43}, 121--135.

\bibitem[\protect\citename{Wang {\em et~al.}\relax, }2013]{wang2013exponential}
Wang, Peng, Pattison, Philippa, \& Robins, Garry. (2013).
\newblock Exponential random graph model specifications for bipartite
  networks---a dependence hierarchy.
\newblock {\em Social networks}, {\bf 35}(2), 211--222.

\bibitem[\protect\citename{Wasserman {\em et~al.}\relax,
  }1994]{wasserman1994social}
Wasserman, Stanley, Faust, Katherine, {\em et~al.\ }\relax. (1994).
\newblock {\em Social network analysis: Methods and applications}.
\newblock Cambridge university press.

\bibitem[\protect\citename{Wuchty {\em et~al.}\relax,
  }2007]{wuchty2007increasing}
Wuchty, Stefan, Jones, Benjamin~F, \& Uzzi, Brian. (2007).
\newblock The increasing dominance of teams in production of knowledge.
\newblock {\em Science}, {\bf 316}(5827), 1036--1039.

\end{thebibliography}

\appendix

\section{Model variants}

In the appendix we report results obtained with additional models that test for additional effects or define explanatory variables in a slightly different way.

\paragraph{Number of past collaborators.}
Subset repetition of order one averages the number of previous events over all participants of a hyperedge. A variation of this statistic is obtained by weighting previous events with their size minus one. The intuition of this modified statistic is that it can make a difference whether actors' past activity happened in collaboration with many others or with few others. Both, subset repetition of order one and the statistic $num.collab$ can be considered as degree effects. Note that in an ordinary graph (with edges connecting exactly two nodes), the degree of a node is the number of incident edges, or equally the number of adjacent nodes. Interestingly, in a hypergraph this equality does not hold since hyperedges can involve a varying number of nodes. Thus, counting the number of adjacent nodes (as for $num.collab$) yields a different statistic than counting the number of incident hyperevents (as for $sub.rep^{(1)}$). Formally, we define the average number of past collaborators of a hyperedge $h$ (weighted by the number of past collaborations) by
\[
num.collab(t,h,G[E;t])=\sum_{v\in h}\sum_{e\in E_{<t}}\chi(v\in h_e)\cdot(|h_e|-1)\cdot\frac{1}{|h|}\enspace.
\]
Similarly, the success-weighted number of past collaborators of $h$ is defined by
\[
num.collab.succ(t,h,G[E;t])=\sum_{v\in h}\sum_{e\in E_{<t}}y_e\cdot\chi(v\in h_e)\cdot(|h_e|-1)\cdot\frac{1}{|h|}\enspace.
\]

\begin{table}
  \caption{Effect of the number of past collaborators and subset repetition of higher order. CoxPH model for publication rates associated with hyperedges. Estimated parameters and standard errors (in brackets). The first model has been already reported in the main text.
    Second model includes the (weighted) number of prior collaborators and the success-weighted number of collaborators. The third model includes subset repetition and prior success up to order 10.}
  \label{tab:assembly_sub-repetition}
  \begin{center}
\begin{tabular}{l@{\quad}r@{\quad}r@{\quad}r}
\hline
 & joint model & (number of collaborators) & (higher order) \\
\hline
$sub.rep^{(1)}$            & $0.317 \; (0.000)^{***}$  & $0.338 \; (0.001)^{***}$  & $0.317 \; (0.000)^{***}$  \\
$sub.rep^{(2)}$            & $-0.033 \; (0.000)^{***}$ & $-0.038 \; (0.000)^{***}$ & $-0.033 \; (0.000)^{***}$ \\
$sub.rep^{(3)}$            & $0.048 \; (0.000)^{***}$  & $0.050 \; (0.000)^{***}$  & $0.039 \; (0.000)^{***}$  \\
$sub.rep^{(4)}$            &                           &                           & $0.016 \; (0.001)^{***}$  \\
$sub.rep^{(5)}$            &                           &                           & $0.050 \; (0.001)^{***}$  \\
$sub.rep^{(6)}$            &                           &                           & $-0.018 \; (0.002)^{***}$ \\
$sub.rep^{(7)}$            &                           &                           & $0.018 \; (0.002)^{***}$  \\
$sub.rep^{(8)}$            &                           &                           & $-0.053 \; (0.003)^{***}$ \\
$sub.rep^{(9)}$            &                           &                           & $0.100 \; (0.004)^{***}$  \\
$sub.rep^{(10)}$           &                           &                           & $-0.079 \; (0.003)^{***}$ \\
$closure$                           & $-0.042 \; (0.001)^{***}$ & $-0.014 \; (0.001)^{***}$ & $-0.072 \; (0.001)^{***}$ \\
$num.collab$                   &                           & $-0.047 \; (0.001)^{***}$ &                           \\
$prior.succ^{(1)}$         & $-0.224 \; (0.002)^{***}$ & $-0.213 \; (0.002)^{***}$ & $-0.212 \; (0.002)^{***}$ \\
$prior.succ^{(2)}$         & $0.025 \; (0.001)^{***}$  & $0.026 \; (0.001)^{***}$  & $0.027 \; (0.001)^{***}$  \\
$prior.succ^{(3)}$         & $0.034 \; (0.001)^{***}$  & $0.033 \; (0.001)^{***}$  & $0.038 \; (0.001)^{***}$  \\
$prior.succ^{(4)}$         &                           &                           & $0.000 \; (0.001)\phantom{^{***}}$        \\
$prior.succ^{(5)}$         &                           &                           & $-0.009 \; (0.001)^{***}$ \\
$prior.succ^{(6)}$         &                           &                           & $0.008 \; (0.001)^{***}$  \\
$prior.succ^{(7)}$         &                           &                           & $0.004 \; (0.001)^{***}$  \\
$prior.succ^{(8)}$         &                           &                           & $0.022 \; (0.001)^{***}$  \\
$prior.succ^{(9)}$         &                           &                           & $-0.025 \; (0.001)^{***}$ \\
$prior.succ^{(10)}$        &                           &                           & $0.005 \; (0.001)^{***}$  \\
$succ.disp$        & $0.086 \; (0.000)^{***}$  & $0.083 \; (0.000)^{***}$  & $0.083 \; (0.000)^{***}$  \\
$num.collab.succ$ &                           & $-0.007 \; (0.001)^{***}$ &                           \\
\hline
AIC                               & 7,882,564.124               & 7,877,465.331               & 7,876,008.586               \\
Num. events                       & 355,977                    & 355,977                    & 355,977                    \\
Num. obs.                         & 3,675,744                   & 3,675,744                   & 3,675,744                   \\
\hline
\multicolumn{4}{l}{\scriptsize{$^{***}p<0.001$, $^{**}p<0.01$, $^*p<0.05$}}
\end{tabular}
  \end{center}
\end{table}

We add these two effects to the RHEM explaining publication rates that has been already reported in the main text, repeated in the first model in Table~\ref{tab:assembly_sub-repetition}, and report its parameters in the second column of that table. We find that the number of past collaborators has a significantly negative effect on the future publication rate of a hyperedge. That is, all other things being equal, if the members of a hyperedge have previously published mostly in large teams then they have a lower publication rate than if they have previously published in smaller teams. This effect even holds for previously successful collaboration (negative effect of $num.collab.succ$). We further observe that the signs and significance levels of all other effects remain unchanged.

\begin{table}
  \caption{Effect of the number of past collaborators and subset repetition of higher order. OLS for impact (normalized number of citations) of published papers. Estimated parameters and standard errors (in brackets). The first model has already been reported in the main text. The second model includes the statistics for the weighted number of past collaborators and the respective statistic weighted by success. The third model includes subset repetition and prior shared success up to order 10.}
  \label{tab:performance_order.10}
  \begin{center}
\begin{tabular}{l@{\quad}r@{\quad}r@{\quad}r}
\hline
 & joint model & (number of collaborators) & (higher order) \\
\hline
(Intercept)                       & $-0.000 \; (0.249)\phantom{^{***}}$       & $-0.000 \; (0.249)\phantom{^{***}}$         & $-0.000 \; (0.249)\phantom{^{***}}$         \\
$sub.rep^{(1)}$            & $-0.504 \; (0.388)\phantom{^{***}}$       & $1.423 \; (0.424)^{***}$    & $-0.600 \; (0.389)\phantom{^{***}}$         \\
$sub.rep^{(2)}$            & $-1.796 \; (0.483)^{***}$ & $-1.911 \; (0.483)^{***}$   & $-1.966 \; (0.485)^{***}$   \\
$sub.rep^{(3)}$            & $-1.959 \; (0.455)^{***}$ & $-0.880 \; (0.464)^{\cdot\phantom{**}}$ & $-0.364 \; (0.692)\phantom{^{***}}$         \\
$sub.rep^{(4)}$            &                           &                             & $0.004 \; (0.835)\phantom{^{***}}$          \\
$sub.rep^{(5)}$            &                           &                             & $-1.099 \; (1.267)\phantom{^{***}}$         \\
$sub.rep^{(6)}$            &                           &                             & $-1.198 \; (1.626)\phantom{^{***}}$         \\
$sub.rep^{(7)}$            &                           &                             & $-3.602 \; (2.023)^{\cdot\phantom{**}}$ \\
$sub.rep^{(8)}$            &                           &                             & $2.770 \; (2.167)\phantom{^{***}}$          \\
$sub.rep^{(9)}$            &                           &                             & $-1.322 \; (2.996)\phantom{^{***}}$         \\
$sub.rep^{(10)}$           &                           &                             & $2.131 \; (2.395)\phantom{^{***}}$          \\
$closure$                           & $-0.228 \; (0.335)\phantom{^{***}}$       & $4.604 \; (0.540)^{***}$    & $0.563 \; (0.369)\phantom{^{***}}$          \\
$num.collab$                   &                           & $-7.895 \; (0.695)^{***}$   &                             \\
$prior.succ^{(1)}$         & $14.610 \; (0.323)^{***}$ & $14.508 \; (0.339)^{***}$   & $14.608 \; (0.323)^{***}$   \\
$prior.succ^{(2)}$         & $4.739 \; (0.417)^{***}$  & $4.669 \; (0.417)^{***}$    & $4.712 \; (0.418)^{***}$    \\
$prior.succ^{(3)}$         & $2.602 \; (0.371)^{***}$  & $2.520 \; (0.371)^{***}$    & $1.992 \; (0.480)^{***}$    \\
$prior.succ^{(4)}$         &                           &                             & $1.341 \; (0.523)^{*\phantom{**}}$      \\
$prior.succ^{(5)}$         &                           &                             & $-1.030 \; (0.516)^{*\phantom{**}}$     \\
$prior.succ^{(6)}$         &                           &                             & $-0.405 \; (0.551)\phantom{^{***}}$         \\
$prior.succ^{(7)}$         &                           &                             & $1.802 \; (0.612)^{**\phantom{*}}$     \\
$prior.succ^{(8)}$         &                           &                             & $-0.933 \; (0.679)\phantom{^{***}}$         \\
$prior.succ^{(9)}$         &                           &                             & $0.503 \; (0.724)\phantom{^{***}}$          \\
$prior.succ^{(10)}$        &                           &                             & $-0.549 \; (0.548)\phantom{^{***}}$         \\
$succ.disp$        & $-3.817 \; (0.284)^{***}$ & $-3.997 \; (0.290)^{***}$   & $-3.834 \; (0.284)^{***}$   \\
$num.auth$               & $8.424 \; (0.271)^{***}$  & $9.982 \; (0.306)^{***}$    & $8.656 \; (0.278)^{***}$    \\
$num.collab.succ$ &                           & $1.003 \; (0.317)^{**\phantom{*}}$     &                             \\
\hline
R$^2$                             & 0.023                     & 0.023                       & 0.023                       \\
Num. obs.                         & 355,977                    & 355,977                      & 355,977                      \\
RMSE                              & 148.666                   & 148.639                     & 148.658                     \\
\hline
\multicolumn{4}{l}{\scriptsize{$^{***}p<0.001$, $^{**}p<0.01$, $^*p<0.05$, $^{+}p<0.1$}}
\end{tabular} 
  \end{center}
\end{table}

We also added these two statistics to RHOM explaining the impact of published papers and report the estimated parameters in the second model Table~\ref{tab:performance_order.10}. We find that the number of past collaborators has a negative effect on the impact of future publications, implying that teams of scientists that have previously co-published in larger teams tend to write papers that attract a lower number of citations -- all other things being equal. (We recall that the number of authors of a paper has a positive effect on its scientific impact and point out that this is not a contradiction.) However, the success-weighted number of past collaborators has a positive effect on future success. Taken together, it seems to be past \emph{unsuccessful} collaborations in large teams that have a decreasing effect on future success.

\paragraph{Subset repetition of higher order.}
In the main text we reported RHEM and RHOM including subset repetition up to order three and prior (shared) success also up to order three. Technically it is possible to include subset repetition (and prior shared success) of higher order -- a hard limit for the order is the maximal size of the observed hyperevents beyond which subset repetition would be constantly zero. We fitted RHEM with subset repetition and prior shared success up to order ten and report the estimated parameters in the third model in Table~\ref{tab:assembly_sub-repetition}. We can observe that the alternating pattern of signs of the subset repetition effect, discussed in the main text, continues. In fact, with the sole exception of $sub.rep^{(4)}$, parameters of successive subset repetition statistics take alternating signs. Prior shared success is mostly positive for publication rates -- although we observe two negative and one non-significant parameters. We also note that the signs and significance levels of all other effects (first model in Table~\ref{tab:assembly_sub-repetition}) are unaffected by the inclusion of subset repetition of higher order. We also included subset repetition and prior shared success of order up to ten in RHOM explaining the impact of published papers (third model in Table~\ref{tab:performance_order.10}). Here we observe that most of the higher order effects are not significant.

\paragraph{Interaction of effects with hyperedge size.}
Models that we reported in the main text assumed that explanatory variables have effects that are \emph{homogeneous} over hyperedges of different sizes. For instance, if RHEM estimated so far revealed a negative closure effect, then it is assumed that closure is negative for small events (scientific papers with few authors) as well as for large events (scientific papers with many authors). This, however, is not necessarily the case: some explanatory variable might have an increasing effect for small events but a decreasing effect for large events. 

\begin{table}
  \caption{Interaction of hyperevent effects with hyperedge size (i.\,e., number of authors of published papers). CoxPH model for publication rates associated with hyperedges. Estimated parameters and standard errors (in brackets). The first model has been already reported in the main text.
    The second model interacts all effects with the size of hyperedges.}
  \label{tab:assembly_size}
  \begin{center}
\begin{tabular}{l@{\quad}r@{\quad}r}
\hline
 & joint model & interacted with hyperedge size \\
\hline
$sub.rep^{(1)}$                         & $0.317 \; (0.000)^{***}$  & $0.487 \; (0.001)^{***}$  \\
$sub.rep^{(2)}$                         & $-0.033 \; (0.000)^{***}$ & $-0.106 \; (0.002)^{***}$ \\
$sub.rep^{(3)}$                         & $0.048 \; (0.000)^{***}$  & $0.103 \; (0.001)^{***}$  \\
$closure$                                        & $-0.042 \; (0.001)^{***}$ & $-1.322 \; (0.014)^{***}$ \\
$prior.succ^{(1)}$                      & $-0.224 \; (0.002)^{***}$ & $-0.222 \; (0.002)^{***}$ \\
$prior.succ^{(2)}$                      & $0.025 \; (0.001)^{***}$  & $0.043 \; (0.001)^{***}$  \\
$prior.succ^{(3)}$                      & $0.034 \; (0.001)^{***}$  & $0.021 \; (0.001)^{***}$  \\
$succ.disp$                     & $0.086 \; (0.000)^{***}$  & $0.064 \; (0.001)^{***}$  \\
$sub.rep^{(1)}$:$num.auth$     &                           & $0.252 \; (0.001)^{***}$  \\
$sub.rep^{(2)}$:$num.auth$     &                           & $-0.158 \; (0.002)^{***}$ \\
$sub.rep^{(3)}$:$num.auth$     &                           & $0.066 \; (0.001)^{***}$                             \\
$closure$:$num.auth$                    &                           & $0.123 \; (0.002)^{***}$                             \\
$prior.succ^{(1)}$:$num.auth$  &                           & $-0.001 \; (0.004)\phantom{^{***}}$                                  \\
$prior.succ^{(2)}$:$num.auth$  &                           & $-0.013 \; (0.001)^{***}$                            \\
$prior.succ^{(3)}$:$num.auth$  &                           & $0.012 \; (0.001)^{***}$                             \\
$succ.disp$:$num.auth$ &                           & $-0.020 \; (0.001)^{***}$                            \\
\hline
AIC                                            & 7,882,564.124               & 7,829,304.742               \\
Num. events                                    & 355,977                    & 355,977                    \\
Num. obs.                                      & 3,675,744                   & 3,675,744                   \\
\hline
\multicolumn{3}{l}{\scriptsize{$^{***}p<0.001$, $^{**}p<0.01$, $^*p<0.05$}}
\end{tabular}
  \end{center}
\end{table}

It is relatively easy to test in RHEM or RHOM for such possible heterogeneity with respect to hyperedge size. In the second model reported in Table~\ref{tab:assembly_size} we interact all effects with the size of hyperedges (number of authors) to test whether we find significant differences between effects for small and for large hyperevents. A first observation is that all baseline effects (i.\,e., without the interaction) keep their signs and significance levels. Some of the interaction effects -- most notably for subset repetition -- go in the same direction as the respective baseline effects. This means that these effects are stronger for large hyperedges (large teams) than for small hyperedges. An interesting exception is given by closure which has a positive interaction effect with hyperedge size. This means that for large groups of scientists (who might or might not co-publish a paper) the effect of closure on co-publication rates is less negative than for smaller groups of actors. Comparing parameter sizes we find that the (negative) closure parameter is about ten times in absolute value the parameter of the interaction effect of closure with hyperedge size. This means that a given hyperedge needs a size that is ten standard deviations above average so that the interaction effect cancels out the negative baseline effect of closure. Beyond that size closure would even have a positive effect on co-publication. The mean number of authors over all observations is 9.6 and the standard deviation is 11.7. This means that in our data there is no single observation for which the number of authors is by 10 standard deviations larger than average (recall that the number of coauthors is bounded by 100 in our data). Thus, we can conclude that closure never has a positive effect in our data -- but its negative effect is less strong for larger hyperedges. 

\begin{table}
  \caption{Interaction of hyperevent effects with hyperedge size (i.\,e., number of authors of published papers). OLS for impact (normalized number of citations) of published papers. Estimated parameters and standard errors (in brackets). The first model has been already reported in the main text.
    The second model interacts all effects with the size of hyperedges. }
  \label{tab:performance_size}
  \begin{center}
\begin{tabular}{l@{\quad}r@{\quad}r}
\hline
 & joint model & interacted with hyperedge size \\
\hline
(Intercept)                                    & $-0.000 \; (0.249)\phantom{^{***}}$       & $0.280 \; (0.282)\phantom{^{***}}$          \\
$sub.rep^{(1)}$                         & $-0.504 \; (0.388)\phantom{^{***}}$       & $-0.064 \; (0.538)\phantom{^{***}}$         \\
$sub.rep^{(2)}$                         & $-1.796 \; (0.483)^{***}$ & $-4.467 \; (0.818)^{***}$   \\
$sub.rep^{(3)}$                         & $-1.959 \; (0.455)^{***}$ & $0.453 \; (0.604)\phantom{^{***}}$          \\
$closure$                                        & $-0.228 \; (0.335)\phantom{^{***}}$       & $11.762 \; (2.885)^{***}$   \\
$prior.succ^{(1)}$                      & $14.610 \; (0.323)^{***}$ & $14.855 \; (0.332)^{***}$   \\
$prior.succ^{(2)}$                      & $4.739 \; (0.417)^{***}$  & $4.215 \; (0.435)^{***}$    \\
$prior.succ^{(3)}$                      & $2.602 \; (0.371)^{***}$  & $2.352 \; (0.383)^{***}$    \\
$succ.disp$                     & $-3.817 \; (0.284)^{***}$ & $-3.918 \; (0.299)^{***}$   \\
$num.auth$                            & $8.424 \; (0.271)^{***}$  & $8.343 \; (0.336)^{***}$                                 \\
$sub.rep^{(1)}$:$num.auth$     &                           & $-0.056 \; (0.821)\phantom{^{***}}$                                      \\
$sub.rep^{(2)}$:$num.auth$     &                           & $-3.911 \; (1.205)^{**\phantom{*}}$                                 \\
$sub.rep^{(3)}$:$num.auth$     &                           & $1.202 \; (0.579)^{*\phantom{**}}$                                   \\
$closure$:$num.auth$                    &                           & $-0.667 \; (0.375)^{\cdot\phantom{**}}$                              \\
$prior.succ^{(1)}$:$num.auth$  &                           & $1.678 \; (0.453)^{***}$                                 \\
$prior.succ^{(2)}$:$num.auth$  &                           & $-1.650 \; (0.629)^{**\phantom{*}}$                                 \\
$prior.succ^{(3)}$:$num.auth$  &                           & $1.915 \; (0.498)^{***}$                                 \\
$succ.disp$:$num.auth$ &                           & $0.966 \; (0.423)^{*\phantom{**}}$                                   \\
\hline
R$^2$                                          & 0.023                     & 0.023                       \\
Num. obs.                                      & 355,977                    & 355,977                      \\
RMSE                                           & 148.666                   & 148.624                     \\
\hline
\multicolumn{3}{l}{\scriptsize{$^{***}p<0.001$, $^{**}p<0.01$, $^*p<0.05$}}
\end{tabular}
  \end{center}
\end{table}

We report in Table~\ref{tab:performance_size} the estimated parameters of RHOM in which we interact all effects with hyperedge size. We observe that many of the baseline effects keep their sign. Some of the interaction effects go in the opposite direction as their respective baseline effect. Most notably prior dyadwise shared success is less positive for the impact of future papers published by larger teams and the prior success disparity of larger teams has a less negative effect on their impact.

\paragraph{Decay in the effect of past events.}

Finally we test whether our empirical findings are affected by a vanishing effect of past events over time. In the models reported so far, explanatory variables based on past events (or their performance) are cumulative in the sense that they add up contributions of past events without considering the difference between the time of the past event to the current time. Previous work in REM \citep{lbsb-mftien-13} and RHEM \citep{lerner2021dynamic} suggested to let the influence of past events decay exponentially over time. So far we had no decay in our explanatory variables since we assumed that in our illustrative empirical setting joint publication events have a rather long term effect into the future. For instance, we assume that if two scientists coauthored a joint paper some five or ten years ago, then this will still have an influence on their collaboration probability today. Likewise, if some scientists have published a paper with exceptionally high impact in the distant past, it might still increase their expected performance today.

\begin{table}
  \caption{Letting the effect of past events decay over time. CoxPH model for publication rates associated with hyperedges. Estimated parameters and standard errors (in brackets). The first model has been already reported in the main text.
    The second model lets the effect of past events on future events decay with a half life of five years.}
  \label{tab:assembly_decay}
  \begin{center}
\begin{tabular}{l@{\quad}r@{\quad}r}
\hline
 & without decay & with decay \\
\hline
$sub.rep^{(1)}$     & $0.317 \; (0.000)^{***}$  & $0.348 \; (0.000)^{***}$  \\
$sub.rep^{(2)}$     & $-0.033 \; (0.000)^{***}$ & $-0.029 \; (0.000)^{***}$ \\
$sub.rep^{(3)}$     & $0.048 \; (0.000)^{***}$  & $0.044 \; (0.000)^{***}$  \\
$closure$                    & $-0.042 \; (0.001)^{***}$ & $-0.056 \; (0.001)^{***}$ \\
$prior.succ^{(1)}$  & $-0.224 \; (0.002)^{***}$ & $-0.222 \; (0.002)^{***}$ \\
$prior.succ^{(2)}$  & $0.025 \; (0.001)^{***}$  & $0.025 \; (0.001)^{***}$  \\
$prior.succ^{(3)}$  & $0.034 \; (0.001)^{***}$  & $0.035 \; (0.001)^{***}$  \\
$succ.disp$ & $0.086 \; (0.000)^{***}$  & $0.082 \; (0.001)^{***}$  \\
\hline
AIC                        & 7,882,564.124               & 7,842,889.948               \\
Num. events                & 355,977                    & 355,977                    \\
Num. obs.                  & 3,675,744                   & 3,675,751                   \\
\hline
\multicolumn{3}{l}{\scriptsize{$^{***}p<0.001$, $^{**}p<0.01$, $^*p<0.05$}}
\end{tabular}
  \end{center}
\end{table}

Yet to test whether a time decay affects our empirical findings we recomputed explanatory variables, and reestimated models, letting the influence of past events decay with a half life of five years, compare \citet{lbsb-mftien-13,lerner2021dynamic}. Parameters for RHEM are reported in Table~\ref{tab:assembly_decay}. We find that there is virtually no change in the sign or significance level of any effect in the model. Some effects become slightly stronger, some become slightly weaker, but the change is relatively small compared to the absolute value of parameters. We do observe, however, that the model with a time decay has a better model fit expressed in a smaller AIC value. Thus, a time decay might be appropriate in coauthor networks.

\begin{table}
  \caption{Letting the effect of past events decay over time. Estimated parameters and standard errors (in brackets). OLS for impact (normalized number of citations) of published papers. The first model has been already reported in the main text.
    The second model lets the effect of past events on future events decay with a half life of five years.}
  \label{tab:performance_decay}
  \begin{center}
\begin{tabular}{l@{\quad}r@{\quad}r}
\hline
 & without decay & with decay \\
\hline
(Intercept)                & $-0.000 \; (0.249)\phantom{^{***}}$       & $-0.000 \; (0.249)\phantom{^{***}}$       \\
$sub.rep^{(1)}$     & $-0.504 \; (0.388)\phantom{^{***}}$       & $-0.960 \; (0.399)^{*\phantom{**}}$   \\
$sub.rep^{(2)}$     & $-1.796 \; (0.483)^{***}$ & $-1.939 \; (0.527)^{***}$ \\
$sub.rep^{(3)}$     & $-1.959 \; (0.455)^{***}$ & $-1.688 \; (0.490)^{***}$ \\
$closure$                    & $-0.228 \; (0.335)\phantom{^{***}}$       & $-0.034 \; (0.345)\phantom{^{***}}$       \\
$prior.succ^{(1)}$  & $14.610 \; (0.323)^{***}$ & $15.046 \; (0.325)^{***}$ \\
$prior.succ^{(2)}$  & $4.739 \; (0.417)^{***}$  & $4.661 \; (0.422)^{***}$  \\
$prior.succ^{(3)}$  & $2.602 \; (0.371)^{***}$  & $2.200 \; (0.375)^{***}$  \\
$succ.disp$ & $-3.817 \; (0.284)^{***}$ & $-3.965 \; (0.278)^{***}$ \\
$num.auth$        & $8.424 \; (0.271)^{***}$  & $8.371 \; (0.272)^{***}$  \\
\hline
R$^2$                      & 0.023                     & 0.023                     \\
Num. obs.                  & 355,977                    & 355,977                    \\
RMSE                       & 148.666                   & 148.630                   \\
\hline
\multicolumn{3}{l}{\scriptsize{$^{***}p<0.001$, $^{**}p<0.01$, $^*p<0.05$}}
\end{tabular}
  \end{center}
\end{table}

We also reestimated RHOM explaining the impact of published papers using explanatory variables with decay and report parameters in Table~\ref{tab:performance_decay}. We also find very little qualitative changes compared to RHOM without decay. The only change in a significance level is that subset repetition of order one becomes slightly negative in the model with decay (before it was also negative, but not significant). The signs and significance levels of all other effects are unchanged. We also can observe a slight improvement in the fit of the model with decay, expressed in a slightly smaller root mean squared error. 

\end{document}